\documentclass[modern]{aastex631}

\usepackage{chemformula, CJK, xspace, soul}

\newcommand{\LCONet}{LCO network\xspace}
\newcommand{\LCOfull}{Las Cumbres Observatory (LCO)\xspace}
\newcommand{\LCO}{LCO\xspace}  
\newcommand{\LOOK}{LOOK Project\xspace}
\newcommand{\LOOKfull}{LCO Outbursting Objects Key (LOOK) Project\xspace}
\newcommand{\NEOx}{NEOexchange\xspace}
\newcommand{\BB}{C/2014 UN$_{271}$ (Bernardinelli-Bernstein)\xspace}
\newcommand{\BBshort}{C/2014 UN$_{271}$ (B-B)\xspace}
\newcommand{\gps}{\ensuremath{g_{\mathrm{P1}}}}
\newcommand{\rps}{\ensuremath{r_{\mathrm{P1}}}}

\turnoffeditone
\turnoffedittwo

\submitjournal{PSJ}

\shorttitle{The LOOK Project: Overview and Year 1 Status}
\shortauthors{Lister et al.}
\graphicspath{{./}{figures/}}

\begin{document}
\begin{CJK*}{UTF8}{gbsn}

\title{The LCO Outbursting Objects Key Project: Overview and Year 1 Status}

\correspondingauthor{Tim Lister}

\author[0000-0002-3818-7769]{Tim Lister}
\affiliation{Las Cumbres Observatory, 6740 Cortona Drive Suite 102, Goleta, CA 93117, USA}
\email{tlister@lco.global}

\author[0000-0002-6702-7676]{Michael S. P. Kelley}
\affiliation{Department of Astronomy, University of Maryland, College Park, MD 20742-0001, USA}
\email{msk@astro.umd.edu}

\author[0000-0002-4043-6445]{Carrie E. Holt}
\affiliation{Department of Astronomy, University of Maryland, College Park, MD 20742-0001, USA}

\author[0000-0001-7225-9271]{Henry H. Hsieh}
\affiliation{Planetary Science Institute, 1700 East Fort Lowell Rd., Suite 106, Tucson, AZ 85719, USA}
\affiliation{Institute of Astronomy and Astrophysics, Academia Sinica, P.O.\ Box 23-141, Taipei 10617, Taiwan}

\author[0000-0003-3257-4490]{Michele T. Bannister}
\affiliation{School of Physical and Chemical Sciences | Te Kura Mat\={u}, University of Canterbury,  Private Bag 4800, Christchurch 8140, New Zealand}

\author[0000-0003-2396-4569]{Aayushi  A. Verma}
\affiliation{School of Physical and Chemical Sciences | Te Kura Mat\={u}, University of Canterbury,  Private Bag 4800, Christchurch 8140, New Zealand}

\author[0000-0002-1105-7980]{Matthew M. Dobson}
\affiliation{Astrophysics Research Centre, School of Mathematics and Physics, Queen's University Belfast, Belfast BT7 1NN, UK}

\author[0000-0003-2781-6897]{Matthew M. Knight}
\affiliation{Department of Physics, U.S. Naval Academy, 572C Holloway Rd., Annapolis, MD, 21402, USA}
\affiliation{Department of Astronomy, University of Maryland, College Park, MD 20742-0001, USA}

\author[0000-0001-9784-6886]{Youssef Moulane}
\affiliation{Department of Physics, Auburn University, Auburn, AL 36849, USA}

\author[[0000-0003-4365-1455]{Megan E. Schwamb}
\affiliation{Astrophysics Research Centre, School of Mathematics and Physics, Queen's University Belfast, Belfast BT7 1NN, UK}

\author[0000-0002-2668-7248]{Dennis Bodewits}
\affiliation{Department of Physics, Auburn University, Auburn, AL 36849, USA}


\author{James Bauer}
\affiliation{Department of Astronomy, University of Maryland, College Park, MD 20742-0001, USA}

\author[0000-0002-1278-5998]{Joseph Chatelain}
\affil{Las Cumbres Observatory, 6740 Cortona Drive Suite 102, Goleta, CA 93117, USA}

\author{Estela Fern{\'a}ndez-Valenzuela}
\affiliation{Florida Space Institute, University of Central Florida, 12354 Research Parkway, Partnership 1 building, Orlando, FL 32828, USA}

\author[0000-0002-9925-0426]{Daniel Gardener}
\affiliation{Institute for Astronomy, University of Edinburgh, Royal Observatory, Edinburgh, EH9 3HJ, UK}

\author{Geza Gyuk}
\affiliation{Adler Planetarium, 1300 S. DuSable Lake Shore Dr., Chicago, IL 60605, USA}

\author{Mark Hammergren}
\affiliation{Farther Horizons LLC}

\author{Ky Huynh}
\affiliation{Department of Astronomy, University of Maryland, College Park, MD 20742-0001, USA}

\author[0000-0001-8923-488X]{Emmanuel Jehin}
\affiliation{Space sciences, Technologies \& Astrophysics Research (STAR) Institute, University of Li\`{e}ge, Li\`{e}ge, Belgium}

\author{Rosita Kokotanekova}
\affiliation{European Southern Observatory (ESO), Karl-Schwarzschild Str. 2, 85748, Garching bei M\"{u}nchen, Germany}
\affiliation{Institute for Astronomy, University of Edinburgh, Royal Observatory, Edinburgh, EH9 3HJ, UK}

\author{Eva Lilly}
\affiliation{Planetary Science Institute, 1700 East Fort Lowell Rd., Suite 106, Tucson, AZ 85719, USA}

\author{Man-To Hui}
\affiliation{Institute for Astronomy, University of Hawaii, 2680 Woodlawn Drive, Honolulu, HI 96822, USA}
\affiliation{Macau University of Science and Technology}

\author[0000-0002-0622-2400]{Adam McKay}
\affiliation{American University/NASA Goddard Space Flight Center, Greenbelt, MD, 20771, USA}

\author[0000-0002-9298-7484]{Cyrielle Opitom}
\affiliation{Institute for Astronomy, University of Edinburgh, Royal Observatory, Edinburgh, EH9 3HJ, UK}

\author[0000-0001-8541-8550]{Silvia Protopapa}
\affiliation{Southwest Research Institute, 1050 Walnut Street, Suite 300, Boulder, CO 80302, USA}

\author{Ryan Ridden-Harper}
\affiliation{School of Physical and Chemical Sciences | Te Kura Mat\={u}, University of Canterbury,  Private Bag 4800, Christchurch 8140, New Zealand}

\author{Charles Schambeau}
\affiliation{Florida Space Institute, University of Central Florida, 12354 Research Parkway, Partnership 1 building, Orlando, FL 32828, USA}

\author[0000-0001-9328-2905]{Colin Snodgrass}
\affiliation{Institute for Astronomy, University of Edinburgh, Royal Observatory, Edinburgh, EH9 3HJ, UK}

\author{Cai Stoddard-Jones}
\affiliation{School of Physics and Astronomy, Cardiff University, Queens Buildings, The Parade, Cardiff CF24 3AA, UK}

\author[0000-0002-8658-5534]{Helen Usher}
\affiliation{The Open University, Walton Hall, Milton Keynes, MK7 6AA, UK}

\author{Kacper Wierzchos}
\affiliation{Catalina Sky Survey, Lunar and Planetary Laboratory, University of Arizona, 1629 E University Blvd, Tucson, AZ 85721-0092}

\author{Padma A. Yanamandra-Fisher}
\affiliation{Space Science Institute, 4765 Walnut St, Suite B, Boulder, CO 80301, USA}

\author[0000-0002-4838-7676]{Quanzhi Ye (叶泉志)}
\affiliation{Department of Astronomy, University of Maryland, College Park, MD 20742-0001, USA}
\collaboration{42}{(The LCO Outbursting Objects Key (LOOK) Project)}

\author[0000-0001-5749-1507]{Edward Gomez}
\affil{Las Cumbres Observatory, School of Physics and Astronomy, Cardiff University, Queens Buildings, The Parade, Cardiff CF24 3AA, UK}

\author[0000-0002-4439-1539]{Sarah Greenstreet}
\affil{Department of Astronomy and the DIRAC Institute, University of Washington, 3910 15th Ave NE, Seattle, WA 98195, USA}



\begin{abstract}
The LCO Outbursting Objects Key (LOOK) Project uses the telescopes of the Las Cumbres Observatory (LCO) Network to: (1) to systematically monitor a sample of \edit1{previously discovered} Dynamically New Comets over the whole sky\edit1{ to assess the evolutionary state of these distant remnants from the early Solar System}, and (2) use alerts from existing sky surveys to rapidly respond to and characterize detected outburst activity in all small bodies. The data gathered on outbursts helps to characterize each outburst's evolution with time, assess the frequency and magnitude distribution of outbursts in general, and contributes to the understanding of outburst processes and volatile distribution in the Solar System.

The LOOK Project exploits the synergy between current and future wide-field surveys such as ZTF, PanSTARRS, and LSST as well as rapid-response telescope networks such as LCO, and serves as an excellent testbed for what will be needed \edit2{for} the much larger number of objects coming from Rubin Observatory.

We will describe the LOOK Project goals, the planning and target selection (including the use of NEOexchange as a Target and Observation Manager or ``TOM"), and results from the first phase of observations, including the detection of activity and outbursts on the giant comet C/2014 UN$_\mathrm{271}$ (Bernardinelli-Bernstein) and the discovery and follow-up of \edit2{28} outbursts on \edit2{14} comets. Within these outburst discoveries, we present a high cadence \edit1{lightcurve} of 7P/Pons-Winnecke with \edit2{10 outbursts observed over 90} days, a large outburst on 57P/duToit-Neujmin-Delporte, and evidence that comet P/2020 X1 (ATLAS) was in outburst when discovered.

\end{abstract}



\section{Introduction}
\label{sec:intro}

\edit1{The Las Cumbres Observatory (LCO) supports several large, coherent, multi-year observing programs
called Key Projects\footnote{\url{https://lco.global/science/keyprojects/}} designed to maximize the science outcomes from its global network of telescopes and their unique capabilities.} The \LOOKfull\footnote{\url{https://www.astro.umd.edu/~msk/science/comae/look/}} is a 3 year Key Project using the many robotic telescopes of the LCO network to study the behavior of active small bodies across the Solar System. Specifically, the LOOK Project has two main objectives:
\begin{enumerate}
    \item Use the telescopes of the LCO Network to systematically image and monitor a sample of Dynamically New Comets (DNCs) \edit1{that have been previously discovered} over the whole sky. These comets are statistically likely to be entering the planetary region from the Oort cloud for the first time. By studying this population's brightness and morphology changes as they pass through the inner Solar System, we can assess their evolutionary state (primitive vs. processed), identify targets for immediate or future follow-up (e.g., outburst vs. ambient coma composition), and, ultimately, better understand the behavior of these distant members as remnants of the early formation of the Solar System. These results will also help optimize the science return of the ESA Comet Interceptor mission \citep{Snodgrass2019} \edit1{by providing a better picture of the evolving activity and behavior of an Oort cloud comet, the intended target for Comet Interceptor} and other future missions as well as improve our understanding of the interstellar objects that are just beginning to be discovered.
    \item Use alerts and other data from the existing sky surveys such as ZTF (Zwicky Transient Facility; \citealt{Bellm2019}), PS1 \& 2, (PanSTARRS; \citealt{Chambers2016}), CSS (Catalina Sky Survey) and ATLAS (Asteroid Terrestrial-impact Last Alert System; \citealt{Tonry2018ATLAS}) to search for outburst activity in small bodies (comets, asteroids, centaurs) and rapidly respond to these outbursts with the telescopes of the LCO Network \citep{Brown2013LCOGT}. These data will characterize each outburst's evolution with time, help assess the frequency and magnitude distribution of outbursts in general, and  contribute to our understanding of outburst processes. This will allow us to gain a better understanding of these outbursts on small bodies and the distribution of volatiles across the Solar System.
\end{enumerate}
The \LOOK's observing program began on 2020 July 1 and will run through 2023 July 31.  It was designed to exploit the synergy between current and future wide-field surveys (such as ZTF, PanSTARRS, ATLAS, and LSST \edit1{(Legacy Survey of Space and Time; \citealt{Ivezic2019})}) and rapid-response telescope networks such as LCO. The techniques, data reduction, and analysis software developed during the \LOOK and demonstrated using the ZTF survey (the main source of \edit1{transient} events) acts as an excellent ``scale model'' and testbed for what will be needed for the much larger number of objects coming from LSST.
Here we describe the LOOK Project goals, the planning and target selection (including the use of \NEOx as a Target and Observation Manager or ``TOM"), and the results from the first year of observations.

\section{Scientific Goals}
\label{sec:goals}

Asteroids and comets are small bodies left over from the formation of the Solar System. Their preserved interiors and exteriors hold key information on the conditions and processes extant at that early epoch and the processes that have shaped the Solar System since then.  Understanding the properties of small bodies in our Solar System is key to understanding our own origins, as well as giving context to our searches for volatiles and the precursors to life in the large number of exoplanetary systems now being discovered.

Although comets have spent billions of years in the cold stability of the outer Solar System, comets do evolve. Their original primitive nature is altered by irradiation, by both solar UV photons, and galactic cosmic rays \citep{Stern2003}. Once in the inner Solar System, heating by the Sun will preferentially cause the loss of the most volatile components (e.g., \citealt{Prialnik2009}), while activity itself also drives both physical changes \citep{Jewitt2004, Veverka2013, Vincent2017} and chemical changes (e.g., \citealt{Huebner2006}; \citealt{Feaga2007}; \citealt{A'Hearn2011}; \citealt{LeRoy2015}) in cometary nuclei. Understanding this evolution is crucial to connecting observed cometary properties back to conditions in the early Solar System, which can then be used to constrain elements of Solar System dynamical evolution models like the streaming instability-based successors (see e.g. \citealt{Morbidelli2020} for a recent review) to the Nice and Grand-Tack models \citep{Levison2009, Walsh2011}. Studies of dynamically new comets and cometary outbursts can help greatly in working towards that understanding.

\edit1{Additionally, }there is a growing understanding that some asteroids in the inner Solar System are active, evolving objects. The relatively recent discovery of active asteroids, including main-belt comets and disrupted asteroids (e.g., the activity of (6478) Gault detected by ATLAS and ZTF; \citealt{Ye2019}), has brought greater interest to this area, and studies of these objects can bring additional insights into the evolution of the different sub-populations of small bodies in the Solar System \citep{Jewitt2015}. Active asteroids appear to be a diverse population driven by a variety of \edit1{potential} mechanisms including volatile sublimation \edit1{\citep{Hsieh2004,Hsieh2011}}, direct impacts \edit1{(e.g. \citealt{Bodewits2011, Jewitt2011})}, and rotation-driven mass shedding \edit1{\citep{Ye2019}}. These mechanisms are central to our understanding of the processes shaping the evolution of asteroids as a whole. Characterizing active asteroids is thus a direct probe of a crucial part of Solar System history.

\subsection{Dynamically New Comets}
\label{sec:DNCs}

The Oort cloud is one of the main reservoirs of cometary objects in our Solar System.  Oort cloud comets with orbital periods $>200$ years are termed long-period comets (LPCs); those of this population with nearly parabolic orbits, which suggest they are entering the inner Solar System for the first time, are referred to as DNCs. This classification can be deduced by examining the original semi-major axis, $a_0$, i.e., the semi-major axis of the comet's orbit before entering the planetary region.  Values of the reciprocal original semi-major axis $1/a_0$ smaller than $4\times10^{-5}$~au$^{-1}$ indicate that the comet has likely not had a previous pass through the inner Solar System  \citep{Dybczynski2015}, and is thus a DNC. More sophisticated orbital integrations and analysis of the Galactic potential and stellar encounters are needed \citep[e.g., ][]{Dybczynski2022} before knowledge of an object's DNC status can be considered secure but the use of the $1/a_0=4\times10^{-5}$~au$^{-1}$ value provides an easy to implement division for observation planning purposes.

The brightness of a comet is a function of the distances from the Sun and observer, but also depends on the amount and type of volatiles and dust in the object. The activity of DNCs at large distances is not well understood. The sublimation of water ice, which typically controls cometary activity closer to the Sun, is inefficient beyond 6 au. Carbon dioxide and carbon monoxide \edit1{\citep{Yang2021}}, more volatile than water and next in order of abundance, are expected to drive activity, including dust emission, at large distances. The small number of comets discovered at $\geq 10$\,au while inbound, and the lack of systematic monitoring, means that the evolution of activity and occurrence of outbursts at greater distances is very uncertain.

\begin{figure}[ht]
\plotone{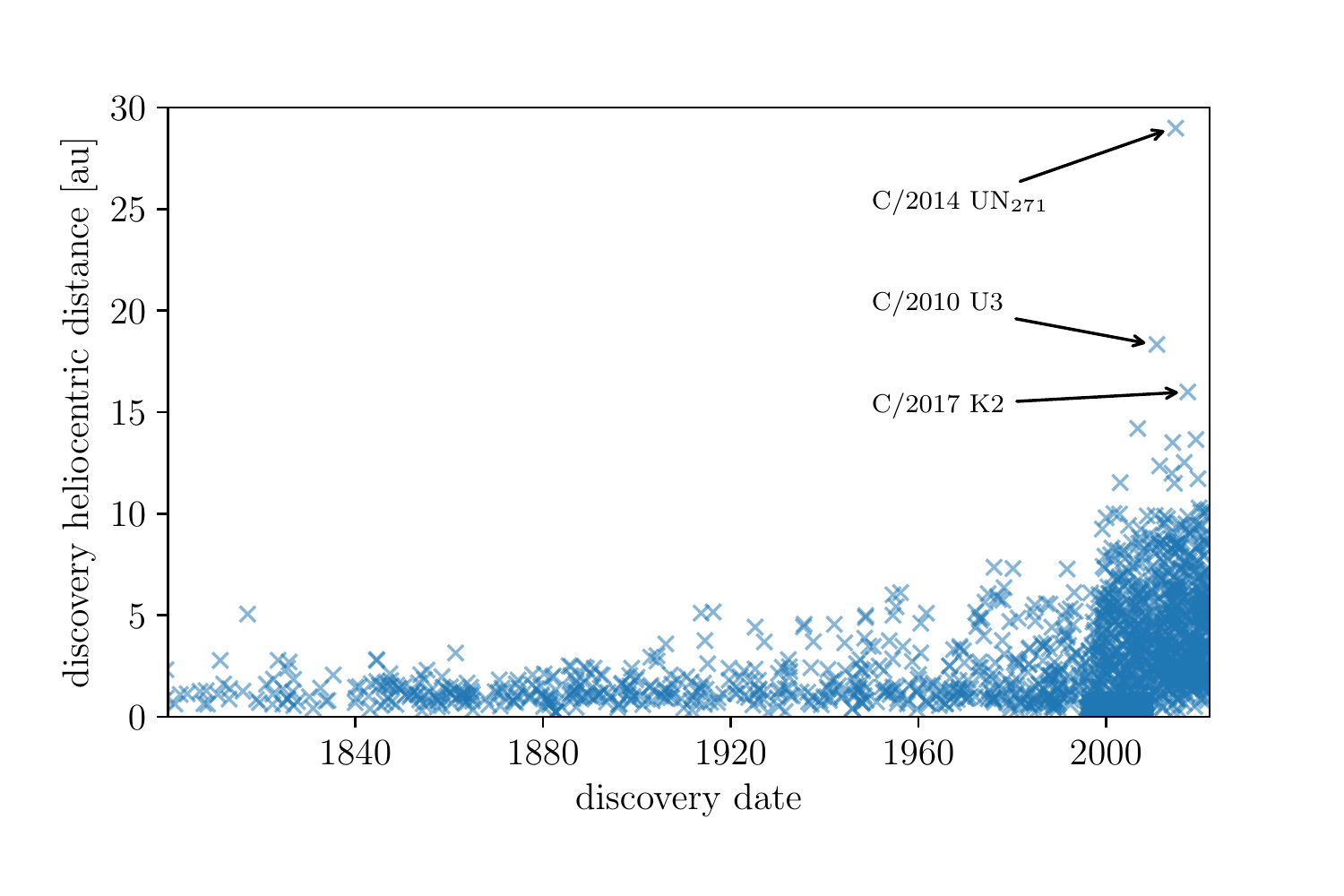}
\caption{Heliocentric distances at the time of discovery of all the long-period comets discovered after 1801 from the MPC database. The density of points increased significantly after 1996, which is when many of the major moving object sky surveys began.
\label{fig:discovery_distances}}
\end{figure}

Surveys of DNCs have been conducted in the past (e.g., \citealt{Whipple1978}; \citealt{Meech2009}) but they have been somewhat limited in either sample size or the range of heliocentric distance that was covered. Current surveys now cover more area to greater depths than ever before, meaning that large numbers of new comets are being discovered and at greater distances from the Sun (Figure~\ref{fig:discovery_distances}). This presents us with an opportunity to greatly improve on past work. Coupled with the availability of flexibly scheduled robotic telescopes such as the \LCONet, conditions are prime for a systematic study of new comets.

Observations of DNCs will also allow us to better understand the processes that drive cometary activity at distance (e.g. by modeling total dust coma brightness; \citealt{Meech2017}) allowing better predictions of the behavior of future comets, including mission targets. This is critical for the ESA Comet Interceptor mission \citep{Snodgrass2019}, which will select its target based on \edit2{the} discovery of a new comet at \edit2{a} large distance, but will only meet it years later as it approaches the Sun; success depends on being able to make reliable predictions of how activity evolves in newly identified objects.

\subsection{Solar System Object Outbursts}
Solar System object outbursts are sudden, short-lived mass-loss events which can span many orders of magnitude in ejection mass and outburst frequency. They vary from small scale outbursts every few rotation periods (seen by the Deep Impact and Rosetta spacecraft at comets 9P/Tempel 1 and 67P/Churyumov-Gerasimenko; e.g., \citealt{Farnham2007}; \citealt{Vincent2016}) to large scale fragmentation or complete disruption events (e.g., \citealt{Jewitt2014}, \citealt{Hughes1990}, \citealt{Ishiguro2016}). While some outbursts may be caused by sublimation of previously covered pockets of ices with low vaporization temperatures, some have been linked to mass shedding due to rotational instability. \edit1{Evidence for mass shedding events on asteroids has been provided by the OSIRIS-REx spacecraft which observed repeated instances of particle ejection \citep{Lauretta2019}.} For asteroids with a diameter $\lesssim1$\,km, spin-up via the YORP effect can occur on timescales as little as a million years (e.g. \citealt{Rossi2009}). As spin rates approach the critical value, sudden structural failure and mass shedding can occur. Even for larger asteroids, small changes in the spin state can rearrange the surface, producing dust events or revealing hidden volatiles \citep{Jacobson2014}.

Distinguishing between types of outbursts and refining physical models require\edit2{s} rapidly detecting outbursts and obtaining time critical measurements of ejected material within hours or days before it disperses, disintegrates or sublimates. Characterizing a statistically meaningful sample of the high amplitude outbursts that can be observed from the ground on large spatial scales will allow connection to the detailed \emph{in situ} measurements made by spacecraft and the rare super outbursts (e.g. \citealt{Li2011}  for the super outburst of 17P/Holmes or \citealt{Ishiguro2016} for an overview), providing greater understanding of Solar System object structure and evolution.

\subsection{LOOK Overview}

\begin{figure}
\begin{center}
\includegraphics[width=\textwidth]{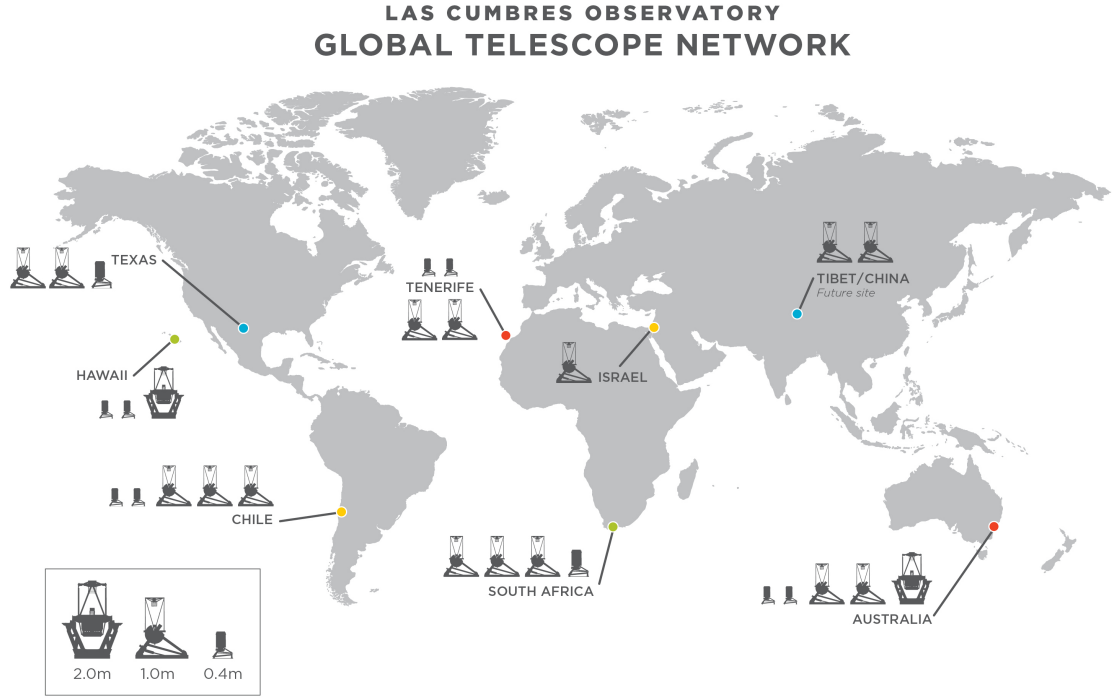}
\end{center}
\caption{Network map of \LCO facilities}
\label{fig:networkmap}
\end{figure}

The \LOOK aims to systematically monitor and characterize a sample of DNCs and obtain rapid response observations of outbursts, improving our understanding of the mechanisms responsible for such outbursts. This is achieved by making use of the \LCO worldwide network of robotic telescopes which is shown schematically in  Figure~\ref{fig:networkmap}. The main telescopes used by \LOOK are the twelve \edit1{identical LCO designed and built} 1-meter telescopes at McDonald Observatory (Texas), Tenerife (Canary Islands, Spain), Cerro Tololo (Chile), SAAO (South Africa) and Siding Spring Observatory (Australia). These 1-meter telescopes join the two existing 2-meter Faulkes Telescopes\edit1{, Faulkes Telescope North (FTN) in Maui, Hawaii and Faulkes Telescope South (FTS) in Australia,} which \LCO has operated since 2005. The whole telescope network has been fully operational since 2014 May, and observations are executed remotely and robotically. Future expansion to a site at Ali Observatory, Tibet with one meter class telescopes (which will not be identical to the existing twelve LCO 1-meter's) in 2022 is also planned.

The 1-meter telescopes have a Sinistro imager consisting of a Fairchild $4096\times4096$ Charge Coupled Device (CCD) and Archon controller, giving a $26\arcmin\times26\arcmin$ field of view (FOV) with 0\farcs389 per pixel and 21 filters. This filter set includes full sets of Bessell ($UBVRI$) and SDSS/PanSTARRS filters ($ugriz_sY$) along with the high throughput PanSTARRS-$w$ (equivalent to $g+r+i$). \LOOK observations are occasionally supplemented by observations with the 2-meter FTN and FTS telescopes through LCO educational partners. The FTN and FTS telescopes originally had identical $10\arcmin\times10\arcmin$ FOV CCD imagers with 18 filters but the imager on the FTN telescope on Maui, HI was replaced in 2020 November with a copy of the MuSCAT2 \citep{Narita2019muscat2} four-channel instrument called MuSCAT3 \citep{Narita2020muscat3}. The MuSCAT3 instrument at FTN makes use of dichroics to provide simultaneous four-color imaging in $g'r'i'z_s$ filters.  Each site also has a single high-resolution ($R \sim$\,53,000) NRES echelle spectrograph \citep{Eastman2014NRES, Siverd2018NRES} with a fiber feed from one or more 1-meter telescopes, but these are not used for \edit1{the} \LOOK. More details of the telescopes and the network are given in \cite{Brown2013LCOGT}.

To assist in scheduling and tracking the \LOOK observations, we make use of a target and observation manager system to aggregate observational information on comets and active objects and coordinate follow-up observations. This will be described in more detail in the following sections.

\section{Target Selection}
\label{sec:targetsel}

The majority of the targets for the DNC monitoring part of the \LOOK come from new comet discoveries reported from the current wide-field sky surveys  (ATLAS, CSS, PS1 \& 2 and ZTF).
Although these surveys are currently all based in the Northern Hemisphere, they cover down to Dec $\geq -30^\circ$, providing plenty of coverage for the greater number of LCO 1-m telescopes in the Southern Hemisphere. More discoveries in the Southern Hemisphere will come from the newly commissioned ATLAS 3 \& 4 telescopes in Chile and South Africa.

Targets are automatically ingested by \NEOx (\citealt{Lister2021NEOx}; see Section~\ref{sec:obssched} for more details) after leaving the Minor Planet Center (MPC)'s NEO or Potential Comet Confirmation pages as confirmed comets. \edit1{In addition, targets within the \NEOx database are annotated with a ``source subtype" based on their orbit type e.g. Jupiter family, Halley type, parabolic, etc. during the ingestion process using information from the JPL Small-Body Database or the MPC orbit database.}
The main discriminator for deciding whether an object is dynamically new is the reciprocal original semi-major axis ($1/a_0$), which is obtained via a query of the Minor Planet Center's orbital database\footnote{\url{https://www.minorplanetcenter.net/db_search/}}. We use a value of $4\times10^{-5}$~au$^{-1}$ as the dividing line for considering whether a comet is a DNC. \edit1{This is used to set a `Dynamically New' source subtype on the target in the database to aid target selection, but does not automatically schedule these targets for observations until after manual review.} As discussed in Section~\ref{sec:DNCs}, this is somewhat simplistic but allows for easy target selection and an immediate start to observations before a more detailed analysis can become available. Additionally, we consider whether the comet was discovered with a heliocentric distance greater than $r_h\gtrsim5$\,au and will be brighter than $V\sim20.5$ for long enough to monitor over at least 1 au. Comets which meet the criteria are scheduled for regular monitoring observations (see Section~\ref{sec:obssched}). We continue to observe all \LOOK LPCs, even after a designation of DNC can be confirmed or ruled out. This gives us a longer baseline to compare DNCs with non-DNCs.

Outbursting objects, both comets and asteroids, are by their nature self-selecting targets. Detection of objects in outburst primarily come from the ZTF survey (see Section~\ref{sec:outburst-comets}) from anomaly detection software (e.g. \textsc{zchecker}; \citealt{kelley2019}). This is achieved by looking for deviations from the normal evolution of the magnitude with time or phase and is typically performed automatically by \textsc{zchecker} but can also be performed by manual examination. Other sources of outbursts are always considered, e.g. in the cases of discoveries by amateur astronomers, or other analyses by the team (see 95P/Chiron in Section~\ref{sec:centaurs}).
In the case of brighter comets, outburst reports often come from the amateur/citizen science community of observers. After assessment of an outburst notification by \LOOK members, including checking for potential stellar contamination within the photometric aperture, follow-up observations to confirm and study the outburst evolution can be quickly scheduled using the \LCONet (see next section) or certain other telescopes (see Section~\ref{sec:mtjohn}).

\section{Observation Planning and Scheduling}
\label{sec:obssched}

The observing program is controlled through an online portal (\NEOx; \citealt{Lister2021NEOx}) which is an example of a Target and Observation Manager (TOM). TOM systems handle the target ingest\edit2{ion} and curation, observation scheduling and recording, and serving any resulting data files for a scientifically-focused investigation such as LOOK. \NEOx was developed to handle NEO candidate follow-up and target characterization and had already been operational for several years \edit2{prior to LOOK}, making it the natural place to handle observation scheduling for LOOK. \NEOx is running in Amazon Web Services (AWS; ``The Cloud") making it available to all the project participants and more reliable (since it's not running on a single server with a single network connection to the outside world).

\begin{figure}[ht]
\plotone{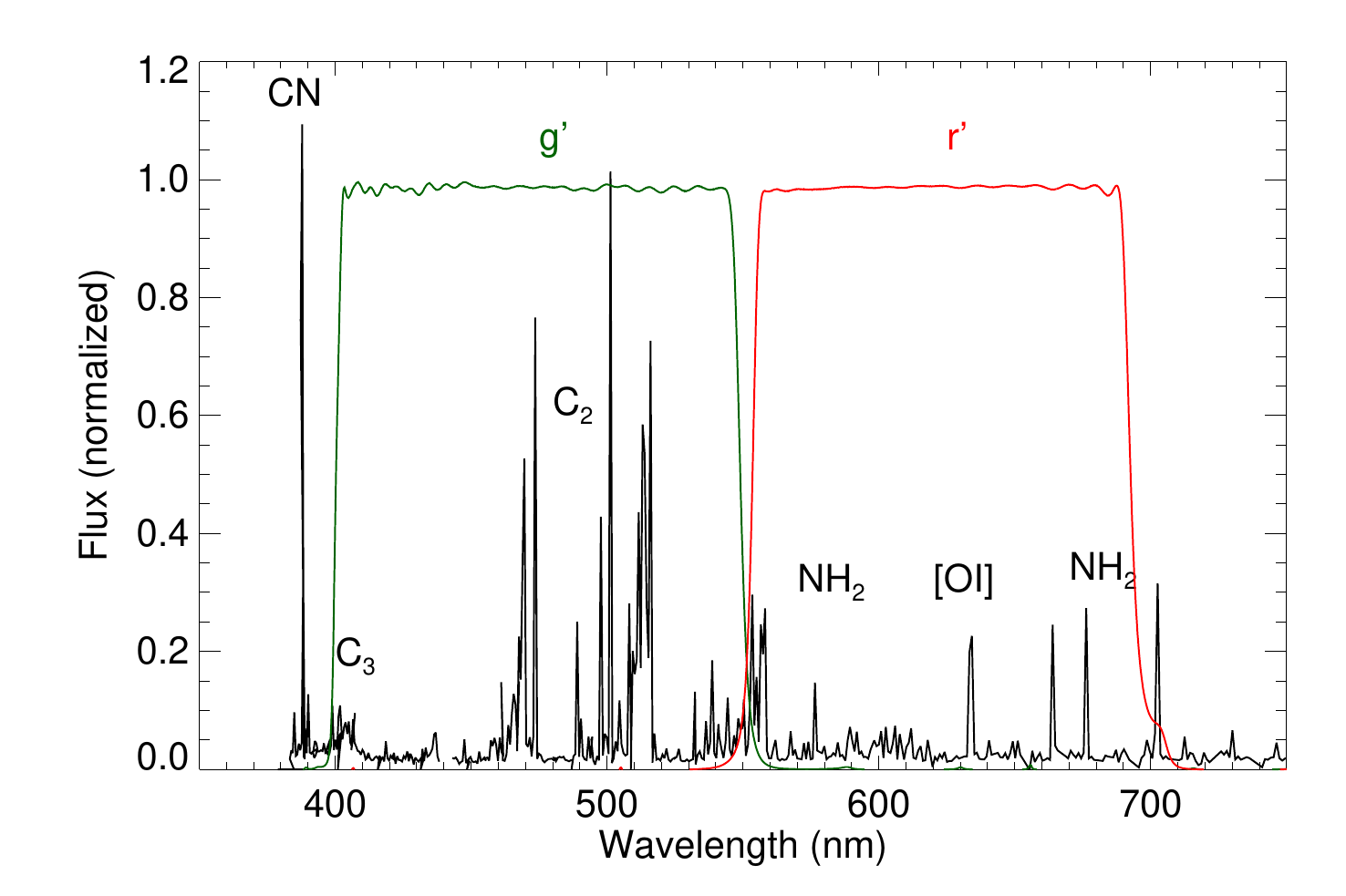}
\caption{Spectrum of 122P/de Vico \citep{Cochran2002} showing cometary gas emission lines and the overlap with the LCO Sloan/PanSTARRS $g'$ and $r'$ filters. The continuum was not removed by \citealt{Cochran2002} as 122P was a very high gas/dust ratio comet and the continuum was negligible compared to the emission lines. The $r'$-filter (red) is free from the strongest cometary gas emission lines, while the $g'$ (green) filter captures many of those. This allows \edit1{for a good estimate of } the different distributions of the gas and dust to be measured separately.
\label{fig:comet_spectrum}}
\end{figure}

The photometric monitoring of DNCs takes place \edit1{every three days} using two filters (Sloan/PanSTARRS $g'$ and $r'$) in order to be sensitive to the gas and dust components of the comet.  In Fig.~\ref{fig:comet_spectrum}, we show a spectrum of the extremely gas-rich comet 122P/de Vico at 0.66~au from the Sun \citep{Cochran2002} and spectral throughputs for the $g'$ and $r'$ filters. The $r'$ band is mostly free of gas emission lines and is thus a good proxy \edit2{for} variations \edit2{in} the amount of dust surrounding the nucleus.  In contrast, the $g'$ filter covers \edit2{the} strong \ch{C_2} and \ch{C_3} emission bands.  Molecules produced in the coma, including \ch{C_2} and \ch{C_3}, tend to have spatial profiles shallower than the dust coma (the latter nominally \edit2{decreases proportional to} distance \edit2{from} the nucleus, $\propto\rho^{-1}$).  Thus, the $g-r$ color of a comet, and its variation with distance to the nucleus $\rho$, can be used to assess the dust-to-gas ratio of the coma.

\NEOx has a LOOK-specific summary page which shows newly discovered comets from the sky surveys (as discussed above in Section~\ref{sec:targetsel}) along with the current monitoring targets. The summary page shows the state of the ongoing monitoring as well as providing a brief view of the observational circumstances (sky position, magnitude, on-sky motion rate and heliocentric distance). The LOOK summary page also provides links to visualization plots which allow the evolution of sky position (in equatorial and galactic coordinates), heliocentric and geocentric distance, magnitude, elongation, and Moon separation, along with the visibility from representative sites in the \LCONet in the Northern and Southern hemispheres.

The LOOK summary page and \NEOx allow monitoring of outburst targets to be quickly scheduled on the \LCONet through a target scheduling form which calculates visibility windows and estimates exposure times based on the brightness and rate of motion, although this is less critical for the slow-moving comets of LOOK than the original NEO targets of \NEOx. LOOK \edit1{DNC} monitoring observations are scheduled using the cadence features of \NEOx and the \LCONet which allows a single observing submission to setup the standard month-long observations cadence of $2\times180$\,s exposures in each of Sloan $g'$ and $r'$ every three days.

For outbursting objects, a key component of the LOOK program and a strength of the robotic, rapid response \LCONet, is to firstly validate these outbursts before triggering Target of Opportunity requests on larger, slower facilities (including \edit2{Hubble Space Telescope}), and secondly characterize the evolution of the post-outburst lightcurve.  Dust moving with radial speeds of 1--100~m~s$^{-1}$ will travel 0.1--10\arcsec{}~day$^{-1}$ for targets a distance of 1~au from the Earth.  Thus, the outburst signal within small photometric apertures decays on day to week timescales.  Gas timescales are faster, given that it nominally moves near 1~km~s$^{-1}$ \citep[e.g.,][]{Tseng2007, Ishiguro2016, Opitom2016}.  Generally, outbursts are at best discovered within 1 to 3 days of onset, and therefore the LOOK Project follow-up observations will best characterize the dust.  However, if a suitable outburst can be detected in a high cadence survey (or part of a survey), there is the possibility that gas can be observed and its dynamics characterized with LCO and LOOK.  Finally, outbursts may be followed by additional events, or by the appearance of fragments (moving $\lesssim1$"/day), motivating observations beyond the initial lightcurve decay.

Follow-up observations of outbursts therefore normally consist of a 10 day-long monitoring with the LCO 1-m telescopes, starting with higher-cadence sampling (every 3 hours) until two days after the outburst discovery, when expected changes are the largest due to devolatilization and dispersion of the dust and the changing photochemistry of the gas. After this initial intensive monitoring, we switch to a slower cadence of once every 8 hours, which is the approximate spacing between the different nodes in the \LCONet (particularly in the Southern hemisphere). The data from the slower cadence is essential for detecting any disintegration and separation of potential fragments, and for characterizing the shape of the light curve, which can be used later to characterize the dynamics and size distribution of the dust ejecta, which leads to an improved estimate of the total mass ejected in the outburst.

Outburst response observations also take place in the Sloan $g'$ and $r'$ filters but with shorter exposure times, typically 60\,s, but more exposures in each filter. The exposure times are shorter for the outburst observations to minimize the \edit2{elongation of the reference stars and less accurate astrometry and photometry} caused by the \edit2{telescope tracking at the} object's \edit2{rate of motion} at each visit. \edit2{This elongation} can be more significant for outbursts than for the DNC monitoring. This is because outbursts typically happen in the inner Solar System when the object's on-sky motion is higher\edit2{ and therefore exposure times need to be shorter to avoid trailing the reference stars}. Observations from LCO Key Projects such as LOOK have a higher priority than the majority of other programs in the LCO scheduler but for responding to outbursts we also have a fraction of  allocated time in ``Time Critical"  mode. This gives a $\sim33\times$ boost in priority when observation requests are being considered by the LCO scheduler. This is used during outburst follow-up to ensure early-time data for the case of an outburst discovered very soon after onset (as discussed above).

\subsection{Follow-up with the Mt John Observatory}
\protect\label{sec:mtjohn}
The 1.8~m telescope at the University of Canterbury's Mt.\ John Observatory (MJUO) in Takap\={o}, New Zealand (1029~m elevation, MPC code 474) has observed since 2020 December in support of the LOOK program. \edit1{MJUO adds additional capacity for LOOK follow-up, along with a certain amount of buffer to bad weather at LCO's Australian site, for targets under daily (or more frequent) monitoring. The additional aperture can  produce higher SNR on fainter targets than the LCO 1m facilities. It also provides a testbed for how classically scheduled and manually operated telescopes can be integrated into the time domain follow-up infrastructure and TOMs.}
Targets are obtained through the \NEOx web portal (\citealt{Lister2021NEOx}) and manually scheduled for in-person observation.
Observations are made with the 2.2 deg$^2$ MOA-cam3 in the MOA-R broadband filter (632--860 nm) \citep{Sako2008}.
For instance, between December 2020 and February 2021, 25 targets, comprising 23 comets and 2 active asteroids, were imaged on 10 separate epochs (Table~\ref{tab:MtJohn}).
The data for these observations are currently under analysis, with MOA-R magnitudes converted to $r_{P1}$ magnitudes via colour corrections implemented in the \texttt{calibrimbore}\footnote{\url{https://github.com/CheerfulUser/calibrimbore}} package.

\begin{deluxetable*}{lrrrrl}
\tablecaption{Targets observed in MOA-R at Mt John Observatory}
\tablecolumns{6}
\tablewidth{0pt} 
\setlength{\tabcolsep}{0.1in}
\renewcommand{\arraystretch}{0.9}
\tablehead{   
  \colhead{Name} & 
  \colhead{$1/a_0$\tablenotemark{$^\alpha$} [au$^{-1}$]}&
  \colhead{N\tablenotemark{$^\beta$}}&
  \colhead{$r_0$\tablenotemark{$^\gamma$} [au]}&
  \colhead{$q$\tablenotemark{$^\delta$} [au]}&
  \colhead{T\edit1{\tablenotemark{$^\epsilon$}[TDB]}}
}
\startdata
189040 & -- & 1 & 1.02 & 0.85 & 2022-01-21 \\
2009 CD & -- & 3 & 1.1 & 0.66 & 2020-07-14 \\
2020 UB5 & -- & 4 & 0.98 & 0.97 & 2019-10-12 \\
2021 CA1 & -- & 3 & 1.02 & 0.99 & 2021-03-05 \\
2021 CD1 & -- & 3 & 1.14 & 0.58 & 2020-07-03 \\
2021 CH & -- & 4 & 1.08 & 0.97 & 2021-03-06 \\
2021 CM & -- & 5 & 1.02 & 1.02 & 2021-01-29 \\
2021 CP & -- & 3 & 0.99 & 0.99 & 2017-09-20 \\
2021 CS & -- & 2 & 1.03 & 0.92 & 2022-01-08 \\
360502 & -- & 1 & 1.05 & 0.70 & 2021-09-10 \\
C/2014 UN271 & 0.000141 & 5 & 20.74 & 10.94 & 2031-01-25 \\
C/2017 W2 & 0.072955 & 2 & 8.4 & 3.95 & 2017-11-01 \\
C/2018 F4 & 0.001391 & 3 & 10.85 & 3.44 & 2019-12-04 \\
C/2019 E3 & 0.000030 & 4 & 11.66 & 10.31 & 2023-11-12 \\
C/2020 R7 & 0.000050 & 3 & 6.24 & 2.96 & 2022-09-16 \\
C/2021 A2 & 0.004282 & 1 & 1.41 & 1.41 & 2021-01-22 \\
C/2021 A6 & 0.000022 & 3 & 7.95 & 7.93 & 2021-05-05 \\
C/2021 A7 & 0.000318 & 6 & 2.86 & 1.97 & 2021-07-15 \\
C/2021 B3 & 0.030214 & 4 & 2.23 & 2.16 & 2021-03-10 \\
C/2021 C5 & 0.000037 & 1 & 7.23 & 3.24 & 2023-02-10 \\
C/2021 G2 & 0.000004 & 1 & 10.55 & 4.98 & 2024-09-10 \\
C/2021 S3 & 0.001919 & 2 & 10.7 & 1.32 & 2024-02-18 \\
C/2022 A2 & -0.000061 & 2 & 7.79 & 1.74 & 2023-02-18 \\
P/2020 V3 & -- & 3 & 6.23 & 6.23 & 2021-02-09 \\
P/2020 V4 & -- & 1 & 5.14 & 5.15 & 2021-07-18 \\
\enddata
\tablecomments{\tablenotemark{$\alpha$} According to MPC database.
\tablenotemark{$\beta$} Number of visits.
\tablenotemark{$\gamma$} Heliocentric distance at the first visit.
\edit1{\tablenotemark{$\delta$} Perihelion distance.}
\edit1{\tablenotemark{$\epsilon$} Time of perihelion.}}
\protect\label{tab:MtJohn}
\end{deluxetable*}

\section{Data processing}
\label{sec:dataproc}

Newly-acquired images from the LOOK Project from the \LCONet are automatically pipeline-processed in near-realtime by the LCO BANZAI (Beautiful Algorithms to Normalize Zillions of Astronomical Images) pipeline  \citep{McCully2018BANZAI}. The BANZAI pipeline performs the standard processing tasks (bias and dark subtraction and flat fielding correction) to produce a basic calibrated data (BCD) product and additionally performs both an astrometric solution and (since 2021 September) photometric zeropoint determination. Processed data products are retrieved from the \edit1{publicly available} LCO Science Archive (which is also located in AWS). These data feed a pipeline at the University of Maryland that performs the domain-specific processing needed for Solar System and extended objects.

The comet outburst pipeline periodically queries the archive for new observations and downloads calibrated data as needed.  A post-download script processes new images one-by-one, including background removal, \edit1{and performs} photometric calibration of the frames with the ATLAS-RefCat2 all-sky photometric catalog \citep{Tonry2018ATLAS} using the \textsc{calviacat} software \citep{Kelley2021-calviacat} and background field stars as measured by \LCO's BANZAI pipeline.  ATLAS-RefCat2 uses the PS1 photometric system, and the comet outburst pipeline considers a color correction owing to the difference between the PS1 filters and the Sloan/PanSTARRS filter set used at \LCO telescopes (the \LCO versions of the Sloan/PanSTARRS filters have higher and more uniform transmission due in part to their smaller physical size; compare the filter profiles in Figure~\ref{fig:comet_spectrum} with Figure~2 of \citealt{Tonry2012PS1}).  Based on LOOK Project photometry, we derived the following color corrections to convert instrumental $g'$ and $r'$ photometry into $g_{\mathrm{P1}}$ and $r_{\mathrm{P1}}$ magnitudes:
  \begin{align}
    g_{\mathrm{P1}} - g'_{\mathrm{inst}} & = - 0.0863\ (g_{\mathrm{P1}} - r_{\mathrm{P1}}) + ZP \\
    r_{\mathrm{P1}} - r'_{\mathrm{inst}} & = +0.0212\ (g_{\mathrm{P1}} - r_{\mathrm{P1}}) + ZP
  \end{align}
where $ZP$ is the best-fit magnitude zero-point for the image.  The $g_{\mathrm{P1}}$ and  $r_{\mathrm{P1}}$ color corrections are based on the analysis of 1008 and 1147 images (standard deviation 0.035 and 0.018~mag), respectively.  Based on these results, we have proceeded with a fixed color correction for all $g'$ and $r'$ images.  The 50- and 90-percentile calibration errors in our initial data set are 0.036 and 0.067~mag for the $g$-band (2175 images), and 0.044 and 0.084~mag for the $r$-band (3228 images).  With the commissioning of additional ATLAS survey facilities in Chile and South Africa, we can hope for a more uniform reference catalog with somewhat greater photometric precision for fainter stars for the southernmost declinations in a later revision of the ATLAS-RefCat without the reprocessing of APASS DR9 survey \citep{Henden2016APASS} data performed for ATLAS-RefCat2 \citep[][their Section 3.3]{Tonry2018refcat2}. This potential non-uniformity in the reference catalog zeropoints in the far southern hemisphere is not a significant source of uncertainty on LOOK comet magnitudes.

Images where the target is missed or the image has other processing issues (missing WCS solution, too few stars for calibration) are automatically saved to a list of files to ignore.  Manual inspection is required to identify some other common issues, such as photometric interference from background sources, or scattered lunar light.  If the image can be calibrated, then photometry is attempted on the comet centroid or the ephemeris position if centroiding fails.  After the target has been measured in the individual frames, data are clustered together by time and telescope\edit1{ (i.e., observing block, median $\Delta t$=870~s)}, and the respective photometry and images are averaged together\edit1{ by filter}.

\section{Results}~\edit1{\footnote{Dates and times of observations in this section are in the UTC time system, except where noted e.g. for times of perihelion which are in TDB.}}
\label{sec:results}

\subsection{DNCs}

The results for the main DNC monitoring part of the LOOK program will be the subject of additional papers (Holt et al. in preparation) but a brief summary of the initial results will be presented here, along with discussion of the special case of the newly-discovered ``extreme'' comet C/2014 UN$_\mathrm{271}$ (Bernardinelli-Bernstein).

\begin{deluxetable*}{lrrrrl}
\tablecaption{Monitored LPCs\label{tab:lookobs}}
\tablecolumns{6}
\tablewidth{0pt} 
\setlength{\tabcolsep}{0.1in}
\renewcommand{\arraystretch}{0.9}
\tablehead{   
  \colhead{Name} & 
  \colhead{$1/a_0$\tablenotemark{$^\alpha$}[au$^{-1}$]}&
  \colhead{N\tablenotemark{$^\beta$}}&
  \colhead{$r_0$\tablenotemark{$^\gamma$}[au]}&
  \colhead{$q$ [au]}&
  \colhead{T\edit1{\tablenotemark{$^\delta$}[TDB]}}
}
\startdata
C/2014 UN$_{271}$ (Bernardinelli-Bernstein)&0.000050&116&20.16&10.95&2031 Jan 22\\
C/2019 F1 (ATLAS-Africano)&-0.000066&110&4.55&3.59&2021 Jun 22\\
C/2019 L3 (ATLAS)&0.000102&57&5.59&3.55&2022 Jan 09\\
C/2020 O2 (Amaral)&-0.000110&70&5.71&4.86&2021 Aug 28\\
C/2020 R7 (ATLAS)&0.000028&90&7.03&2.95&2022 Sep 16\\
C/2020 U4 (PANSTARRS)&0.000021&52&6.57&5.35&2022 Apr 07\\
C/2021 A1 (Leonard)&0.000375&13&3.55&0.61&2022 Jan 03\\
C/2021 C5 (PANSTARRS)&0.000037&30&6.78&3.24&2023 Feb 10\\
C/2021 E3 (ZTF)&0.000033&69&4.82&1.77&2022 Jun 11\\
C/2021 G2 (ATLAS)&0.000004&50&9.44&4.97&2024 Sep 09\\
C/2021 O3 (PANSTARRS)&0.000053&50&4.20&0.29&2022 Apr 20\\
C/2021 P4 (ATLAS)&0.003461&39&4.07&1.08&2022 Jul 31\\
C/2021 Q4 (Fuls)&0.000147&29&8.46&7.56&2023 Jun 08\\
C/2021 S3 (PANSTARRS)&0.001919&21&8.60&1.33&2024 Feb 18\\
C/2021 T4 (Lemmon)&-0.000202&21&8.57&1.48&2023 Jul 31\\
\enddata
\tablecomments{\tablenotemark{$\alpha$} According to MPC database.
\tablenotemark{$\beta$} Number of \edit1{completed} visits \edit1{(through 2022 Feb 21)}.
\tablenotemark{$\gamma$} Heliocentric distance at the first visit.
\edit1{\tablenotemark{$\delta$} Time of perihelion.}}
\protect\label{tab:LPCmonitor}
\end{deluxetable*}

\begin{figure}
    \centering
    \includegraphics[width=0.48\textwidth]{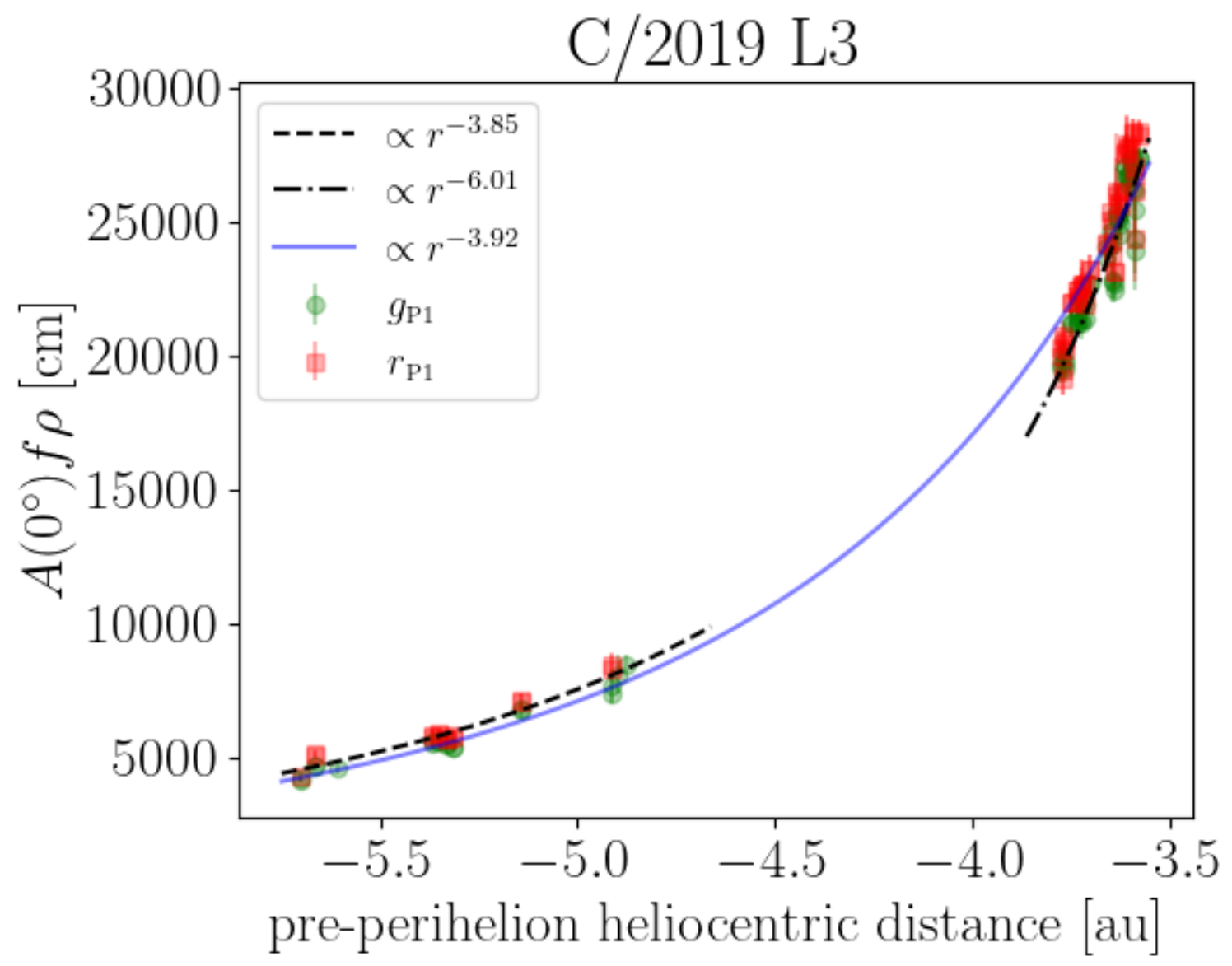}
    \includegraphics[width=0.48\textwidth]{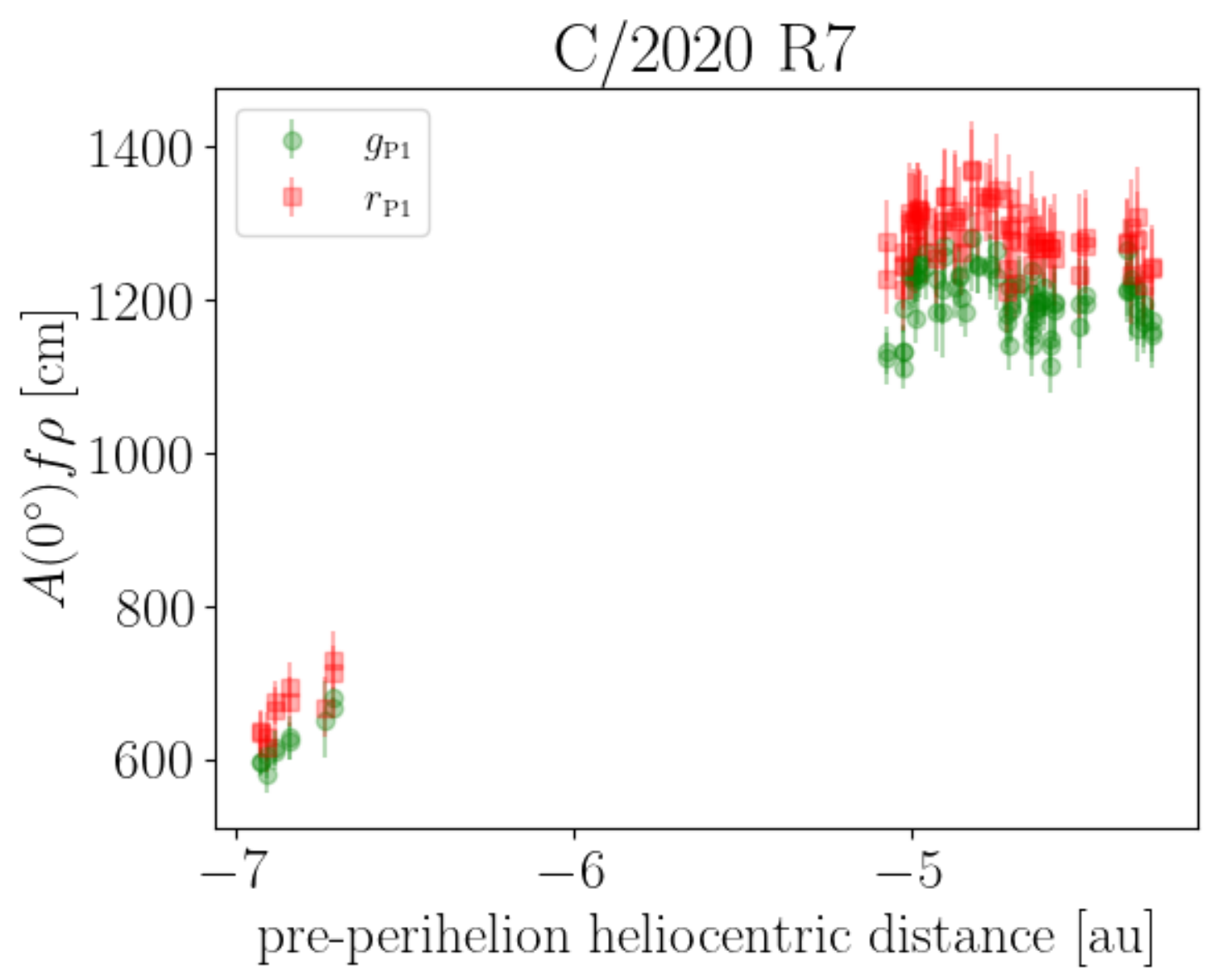}

    \caption{$A(0^\circ)f\rho$ values versus heliocentric distance ($<0$ for pre-perihelion observations) for returning comet, C/2019 L3 (left), and dynamically new comet, C/2020 R7 (right), using photometry from LOOK with 5\arcsec{} radius apertures and the $g'$ and $r'$ filters. C/2019 L3 shows a power-law increase in activity (proportional to $r_h^{-3.9}$ overall) whereas C/2020 R7's activity plateaued after an initial increase. Such differences are expected for returning versus dynamically new comets \citep[e.g.,][]{A'Hearn1995}.}
    \label{fig:DNC_example}
\end{figure}

In the first \edit1{$\sim16$} months, LOOK has monitored 15 LPCs with an average dataset spanning 265 days (see summary in Table~\ref{tab:LPCmonitor}). Of the 15 LPCs, \edit2{eight} currently have $1/a_0<4 \times 10^{-5}$~au$^{-1}$, suggesting they are probably dynamically new. \edit1{These 15 \edit2{LPCs} have been observed at a broad range (2--12) of the possible 1-m telescopes, with a small spike (4 \edit2{LPCs}) observed at the eight 1-m telescopes in the Southern Hemisphere and a smaller spike (2 \edit2{LPCs}) observed at the same eight telescopes plus the two on Tenerife, Canary Islands.}  To highlight some early results, beginning in the second half of 2020, we have been monitoring returning comet C/2019 L3 (ATLAS) and the dynamically new comet C/2020 R7 (ATLAS) (Fig. \ref{fig:DNC_example}). Using $Af\rho$, an aperture-independent quantity used as a proxy for dust production \citep{A'Hearn1984}, we find these comets exhibit different brightness behaviors as they approach perihelion.  $Af\rho$ is the product of the albedo ``$A$", the filling factor ``$f$" (i.e., how much the total cross-section of grains fills the field-of-view), and the radius of the \edit2{photometric aperture} ``$\rho$". C/2019 L3 exhibited a power-law increase in activity with decreasing heliocentric distance, typical of returning comets, while C/2020 R7 first exhibited an activity increase which plateaued as the comet crossed 5 au. Dynamically new comets are known to have a much shallower increase in activity \citep[e.g.,][]{A'Hearn1995}. However, the details of the brightness behavior, particularly at large heliocentric distances, are not well characterized or understood.

As examples of observations to come, we have recently begun monitoring LPCs C/2021 S3 (PanSTARRS) and C/2021 T4 (Lemmon) starting with heliocentric distances of $r_h\gtrsim8.5$\,au that will eventually reach perihelion below 1.5 au. Currently, C/2021 T4 is designated as dynamically new while C/2021 S3 is not. With LOOK, we will be able to directly compare the brightness behavior of these objects as they move into the region where water sublimation becomes an efficient driver of activity.

\subsection{C/2014 UN$_\mathrm{271}$ (Bernadinelli-Bernstein)}
One early result from LOOK was the discovery of cometary activity (\citealt{Kokotanekova2021}, see also \citealt{Buzzi2021}) around the trans-Neptunian object 2014 UN$_{271}$ at $r_h$=20~au within 24 hours of the publication of the initial Dark Energy Survey discovery in MPEC 2021-M53 \citep{Bernardinelli2021MPEC} on 2021 June 19. Later analysis of prior TESS data of the now designated comet \BB indicated a large coma in 2018 at 23~au \citep{Farnham2021} and no detectable rotation period \citep{Ridden-Harper2021}. This comet is unique because of the distance at which it was discovered (as distant as 29~au; \citealt{Bernardinelli2021MPEC}) and for its exceptionally large size.
Initially estimated to be 75~km in radius \citep{Bernardinelli2021}, \cite{Lellouch2022} recently reported a surface-equivalent diameter of $137\pm17$\,km and a geometric albedo in $R$ band of $p_R=5.3\pm1.2\%$ using ALMA.
This confirms that C/2014 UN$_\mathrm{271}$ is the largest Oort-cloud origin object ever found by far, being almost twice as large as comet C/1995 O1 Hale-Bopp, and except for the outbursting Centaur 95P/Chiron (Section~\ref{sec:centaurs}), the largest known comet in the Solar System.

LOOK has been monitoring \BB with the \LCONet since its announcement, along with DDT (Director's Discretionary Time) imaging and spectroscopic observations using larger telescopes, which will be the subject of later publications \citep{Kokotanekova2022}. During this time we have detected an apparent outburst (\citealt{Kelley2021-BB}; Table~\ref{tab:outbursts}) in 2021 September which will be analyzed in more detail in a forthcoming publication (Kelley et al. 2022, \edit1{submitted}).

\subsection{Outbursts across the Solar System}

\subsubsection{Active Asteroids}
\label{sec:outburst-ast}

Asteroid (248370) 2005 QN$_{173}$, now also designated as Comet 433P/(248370) 2005 QN$_{173}$, was discovered to be active on 2021 July 7 by the ATLAS survey \citep{fitzsimmons2021_2005QN173}.  Given subsequently identified archival observations from the Dark Energy Camera on the Blanco Telescope \citep{chandler2021_2005qn173} showing that 433P was active during a previous perihelion passage, this object belongs to the subset of active asteroids known as main-belt comets, whose activity is \edit1{associated with} the sublimation of volatile ices \citep[e.g.,][]{snodgrass2017}. The LOOK Project observations using LCO's network of 1-m telescopes that began on 2021 July 10 form the core of one of the first papers to be published about this object \citep{Hsieh2021_433P}.  Here, densely sampled LOOK data (10 visits within the first 38 days after the initial detection of activity) were crucial for tracking the evolution of the comet's near-nucleus coma as well as its dust tail.  These data showed that while the coma was fading over the period of observations reported in the paper (2021 July 9 to 2021 August 14\edit1{; perihelion was 2021 May 13}), the tail's surface brightness remained roughly constant, indicating that the tail likely contained larger (and thus longer-lived) particles on average than the coma.

Using additional data obtained \edit2{by} LCO's Faulkes Telescope North, the Lowell Discovery Telescope, and the Palomar Hale Telescope, \citet{Hsieh2021_433P} also found that the comet's near-nucleus region and dust tail had similar broadband colors, which implies that no significant gas coma was present. Furthermore, the terminal velocities of ejected dust particles appeared to be extremely low (based on the extremely narrow width of the tail as measured perpendicular to the object's orbit plane), suggesting that the observed dust emission may be aided by rapid rotation of the object's nucleus.  LOOK observations of this object continued until 2022 January 25 
(when the object reached an orbital true anomaly of $\nu\sim70^{\circ}$), giving the prospect of potentially being able to place strong constraints on the turn-off point of the activity, which has not been previously achieved for a main-belt comet.

\subsubsection{Comet Outbursts}
\label{sec:outburst-comets}

Within the first $\sim16$ months of operation of LOOK we have discovered and/or followed up on \edit2{28} outbursts of \edit2{14} comets.  A summary of all events is given in Table~\ref{tab:outbursts}.  The outburst strengths are expressed as change in apparent brightness with respect to a previous measurement or the lightcurve extrapolated to the outburst observation.

\begin{deluxetable}{lcccl}
    \tablecaption{LOOK Project observed outbursts.}
    \tablewidth{0pt}
    \tabletypesize{\footnotesize}
    \tablehead{
      & \colhead{\edit1{Discovery} Date}
      & \colhead{$\Delta m$\tablenotemark{a}}
      & \colhead{$\rho$\tablenotemark{b}}
      \\
      \colhead{Comet}
      & \colhead{(UTC)}
      & \colhead{(mag)}
      & \colhead{(arcsec)}
      & Source
    }
    \startdata
    \multicolumn{5}{c}{Discovery and follow-up} \\\hline
    7P/Pons-Winnecke & 2021 Mar 19.465 & $-0.23\pm0.02$ & 5 & ZTF; \citealt{Kelley2021-7P-a} \\
    & 2021 Jun 02.06 & $-1.21\pm0.06$ & 22 & Amateur and ZTF; \citealt{vanBuitenen2021-7P}, \citealt{Kelley2021-7P-b} \\
    & 2021 Jun 05.25 & $-0.18\pm0.03$ & 5 & LOOK; this work \\
    & 2021 Jun 09.60 & $-0.14\pm0.03$ & 5 & LOOK; this work \\
    & 2021 Jun 15.42 & $-0.34\pm0.04$ & 5 & LOOK; this work \\
    & 2021 Jun 22.08 & $-0.45\pm0.04$ & 5 & LOOK; this work \\
    & 2021 Jul 02.61 & $-0.23\pm0.07$ & 5 & LOOK; this work \\
    & 2021 Jul 10.11 & $-1.10\pm0.06$ & 5 & LOOK; this work \\
    & 2021 Jul 18.93 & $-1.14\pm0.07$ & 5 & LOOK; this work \\
    & \edit2{2021 Aug 25.}\edit2{64} & \edit2{$-0.17\pm0.09$}\tablenotemark{c} & \edit2{5} & \edit2{LOOK; this work} \\
    22P/Kopff & 2021 Apr 19.21 & $-1.12\pm0.17$ & 5.7 & ZTF; \citealt{Kelley2021-22P} \\
    29P/Schwassmann-Wachmann 1 & 2021 Oct 16.88 & $-0.21\pm0.06$ & 5 & MISSION 29P; \citealt{Sharma2021-29P} \\
    67P/Churyumov-Gerasimenko & 2021 Nov 17.86 & $-0.64\pm0.08$ & 5 & ZTF; \citealt{Kelley2021-67P} \\
    97P/Metcalf-Brewington & 2021 Nov 02.27 & $-1.5\pm0.1$ & 15 & ZTF; \citealt{Kelley2021-97P} \\
    191P/McNaught & 2021 Nov 18.81 & $-1.64\pm0.11$ & 5 & ZTF; \citealt{Kelley2021-191P} \\
    382P/Larson & 2021 Sep 27.23 & $-0.98\pm0.09$ & 5 & ZTF; \citealt{Kelley2021-382P} \\
    C/2020 R4 (ATLAS) & 2021 Apr 24.30 & $-1.03\pm0.03$ & 14 & ZTF; \citealt{Kelley2021-2020R4-a} \\
    & 2021 Apr 27.47 & $-1.14\pm0.05$ & 14 & ZTF; \citealt{Kelley2021-2020R4-a} \\
    & 2021 May 06.21 & $-0.60\pm0.05$ & 14 & ZTF; \citealt{Kelley2021-2020R4-a} \\
    & 2021 Jun 04.22 & $-1.39\pm0.07$ &  7 & ZTF; \citealt{Kelley2021-2020R4-b} \\
    \BBshort & 2021 Sep 09.92 & $-0.65\pm0.06$ & 4 & LOOK; \citealt{Kelley2021-BB} \\\hline
    \multicolumn{5}{c}{Follow-up only} \\\hline
    29P/Schwassmann-Wachmann 1 & 2020 Sep 25.34 & $-5.7$\tablenotemark{d} & 5.2 & MISSION 29P; \href{https://britastro.org/node/25120}{R.~Miles, report } \\
    44P/Reinmuth 2 & 2021 Jul 13.19 & $-1.1\pm0.1$ & 5 & ZTF; \citealt{Kelley2021-44P} \\
    57P/duToit-Neujmin-Delporte & 2021 Oct 17.74 & $\sim-3$ & 5.8 &Amateur; \href{https://groups.io/g/comets-ml/message/30140}{comets-ml msg \#30140} \\
    & 2021 Oct 30    & $\sim-0.9$ & 5  &  \edit2{Amateur; \href{https://groups.io/g/comets-ml/message/30180}{comets-ml msg \#30180}; LOOK\tablenotemark{e}} \\
    99P/Kowal 1 & 2021 May 14.18 & $-0.72\pm0.06$ & 5 & ZTF; \citealt{Kelley2021-99P} \\
    120P/Mueller 1 & 2021 Aug 08.48 & $-1.4\pm0.3$ & 7 & ZTF; \citealt{Kelley2021-120P} \\
    P/2020 X1 (ATLAS) & 2020 Dec 01.46 & $<-2$ & & LOOK; \citealt{Fitzsimmons2020-P2020X1}; this work\tablenotemark{f} \\
    \enddata
    \tablecomments{\tablenotemark{a} Outburst strength measured with respect to the ambient coma.  This quantity is aperture dependent.  \tablenotemark{b} Photometric aperture radius for $\Delta m$. \tablenotemark{c} \edit1{The variation in brightness is not \edit2{statistically} significant in this metric, but the outburst is confirmed by inspection of the morphology.}  \tablenotemark{d} Series of outbursts or an extended brightening culminating in the given outburst strength. \tablenotemark{e} This was independently announced first as a $\sim-0.4$ outburst on 2021 Nov 06.72 in \href{https://groups.io/g/comets-ml/message/30180}{comets-ml msg \#30180}  \tablenotemark{f} Although not identified as an outburst of P/2020 X1 (ATLAS) at the time of discovery by \citet{Fitzsimmons2020-P2020X1}, their data suggested an outburst, which is confirmed by our follow-up observations.}
    \protect\label{tab:outbursts}
\end{deluxetable}

Observed outburst strengths range from $-0.14$ to $-5.7$~mag.  For the smallest events, it is difficult to distinguish between outbursts and normal variability, except that a periodicity would be expected in the latter.  Regardless, the LOOK Project data set demonstrates that small transient events may be readily discovered and studied in cometary comae with the \LCONet.  Analysis of the source of the outburst's detection in Table~\ref{tab:outbursts} shows that the majority ($14/28\sim50\%$) have come from the ZTF survey (with LOOK confirmation in the vast majority of cases) with LOOK responsible for $10/28\sim36\%$ and amateur/semi-professionals contributing the remainder ($5/28\sim17\%$). \edit2{(We consider the 2021 Jun 02.06 outburst of 7P to have been jointly discovered by both amateurs and ZTF \citep{vanBuitenen2021-7P,Kelley2021-7P-b} and include it in both counts}).

Below, we present a high-cadence study of comet 7P/Pons-Winnecke, a follow-up study of a large outburst of comet 57P/du~Toit-Neujmin-Delporte, and evidence that comet P/2020~X1 (ATLAS) was in outburst at the time of its discovery.

\paragraph{7P/Pons-Winnecke}
\label{sec:7Poutburst}
Comet 7P is a Jupiter-family comet (JFC) in a 6.32 year orbit with a perihelion distance of 1.23 au. It was discovered on 1819 June 12, and was subsequently lost, rediscovered, and observed numerous times at its close approaches \citep{Kronk2003-Cometography-Vol2}.  7P reached its most recent perihelion on 2021 May 27.  A small, --0.23$\pm$0.02~mag outburst of the comet was observed with ZTF on 2021 March 19.465 and confirmed with prompt LOOK follow-up observations \citep{Kelley2021-7P-a}.  Following the discovery of a larger, --1.21$\pm$0.06~mag event on 2021 June 02.06 \citep{vanBuitenen2021-7P,Kelley2021-7P-b}, we initiated our nominal outburst follow-up observation sequence, during which another outburst occurred.  With three confirmed outbursts of this comet and the target moving into the southern hemisphere (outside the typical NEO survey regions), we continued with an intensive outburst discovery campaign until 2021 August 31 with a median cadence of 14~hr (full range: 1.3~hr to 4~days). Some additional data on 2021 June 05, 09 \& 14 were obtained by Helen Usher and Richard Miles using the LCO 2-m FTN+MuSCAT3 instrument.  We also obtained sparse photometry of the comet with the TRAPPIST telescopes \citep{Jehin2011}.

\begin{figure}
    \centering
    \includegraphics{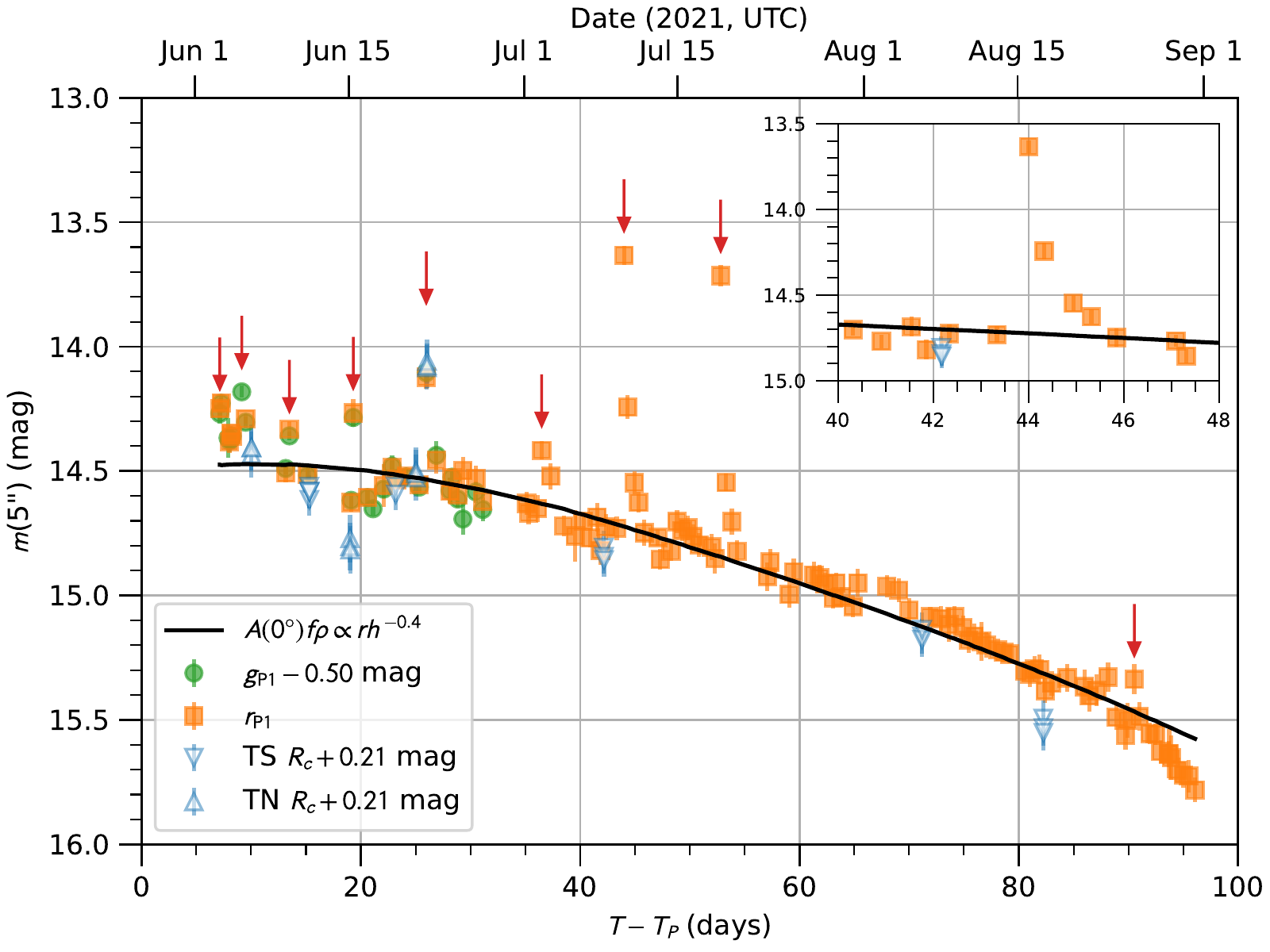}
    \caption{LOOK Project and TRAPPIST lightcurve of comet 7P/Pons-Winnecke, measured within 5\arcsec{} radius apertures. \edit1{This plot and the subsequent three show the time on the lower x-axis as a function of time from perihelion.}  The \gps-band photometry has been offset by the measured color of the inner coma, $\gps-\rps=0.50\pm0.01$~mag, to produce effective \rps-band data.  TRAPPIST South (TS) and TRAPPIST North (TN) $R_c$-band photometry has been offset by the color of the Sun: $R_c-\rps=-0.21$~mag \citep{Willmer2018}.  An approximate secular trend-line is plotted, based on an $Af\rho$ model: $A(0\degr)f\rho=148\ r_h^{-0.4}$~cm.  Arrows mark outbursts of the comet.  An inset shows the lightcurve of the 2021 July 10 outburst in detail.}
    \label{fig:pons-winnecke}
\end{figure}

Our assembled post-perihelion lightcurve based on 5\arcsec{} radius aperture photometry is presented in Fig.~\ref{fig:pons-winnecke}, including supporting photometry from the TRAPPIST telescopes. More details about TRAPPIST data reduction and image processing are given in \citet[]{Moulane2018} and references therein.  Within our lightcurve, \edit2{ten} outbursts have been identified (Table~\ref{tab:outbursts}\edit2{; the first 7P outburst was pre-perihelion and therefore does not appear in Fig.~\ref{fig:pons-winnecke}}).  All events were confirmed based on a visual inspection of the images after subtracting a reference image of the comet either before or after the event.  Including an additional --0.29$\pm$0.07~mag outburst discovered on 2021 June 7.858 by \citet{Sharma2021-7p}\edit2{, and not observed by LOOK, brings the total to ten outbursts in the 90~day period post-perihelion. Including the pre-perihelion outburst listed in Table~\ref{tab:outbursts}}, and a tentative --0.38$\pm$0.12~mag outburst discovered 2021 February 03.446 by \citet{Kelley2021-7P-a}, this comet had at least \edit1{12} outbursts over a 183~day period.

\edit1{Aside from the prolific comet 29P/Schwassmann-Wachmann~1 \citep{Trigo-Rodriguez2008}, no other comet has shown evidence for so many outbursts from ground-based data sets.  For comet 7P/Pons-Winnecke, the observed outburst frequency is at least partially a consequence of the observational circumstances.  \citet{Kelley2021-46P} proposed an outburst discoverability metric, $D$, that assesses the difficulty to detect outbursts given a comet's ambient coma brightness.  The motivating argument is that small aperture sizes limit the amount that the ambient coma (an extended source) can dilute the signal from the outburst (which is initially point-source like).  Thus, the metric is inversely proportional to mass-loss rate (the higher the mass-loss rate, the more difficult it is to detect an outburst of a given size) and comet-observer distance (the closer the comet, the smaller the aperture projected at the distance of the comet).  With the mass-loss rate parameterized by the quantity $Af\rho$, the discoverability metric is:}
\begin{equation}
    D \propto \frac{1}{Af\rho\ \Delta}.
\end{equation}
\edit1{We fit the quiescent $Af\rho$ values for comet 7P with a function proportional to $r_h^k$, and find $A(0\degr)f\rho = 148 r_h^{-0.4}$~cm (plotted in Fig.~\ref{fig:pons-winnecke}).  Therefore, the outburst discoverability metric varies from $\sim0.02$ to 0.05~cm$^{-1}$~au$^{-1}$ in our data set (median 0.04~cm$^{-1}$~au$^{-1}$).  \citet{Kelley2021-46P} found six outbursts of comet 46P in their lightcurve of that comet taken during its historic close approach to Earth (perigee=0.077~au).  The $D$ values for 46P ranged from 0.01 to 0.2~cm$^{-1}$~au$^{-1}$ (median 0.08~cm$^{-1}$~au$^{-1}$), and \citet{Kelley2021-46P} showed that a more typical short period comet would have values at least 10 times smaller.  Comet 46P was already an outlier for the number of outbursts seen at a single comet in telescopic data \citep[cf.][]{Ishiguro2016}, but given that the outburst discoverability for 46P was more favorable than for 7P, we conclude that the frequency of outbursts at 7P is at least four times larger than observed at 46P (0.066~day$^{-1}$ vs 0.016~day$^{-1}$).}

\paragraph{57P/duToit-Neujmin-Delporte}
\label{sec:57Ppoutburst}
An apparent outburst of 57P/duToit-Neujmin-Delporte was reported\footnote{\url{https://groups.io/g/comets-ml/message/30140}} on the \textit{comets-ml} mailing list by F. Kugel. They reported an apparent brightening of $\sim3$\,mag based on observations obtained with a 0.4-m telescope on 2021 October 17.74.  Investigation of the MPC database constrained the outburst to have occurred between two ATLAS observations on 2021 October 5 and October 14. LOOK follow-up observations started on 2021 October 19.77 show the initial decline from the outburst. LOOK and TRAPPIST South observations (which began on 2021 Nov 3) show evidence for an additional outburst around 2021 Nov 2 (an apparent outburst of $\sim1$\,mag. in the LOOK data on 2021 October 29 is due to stellar contamination with a bright $G=12.9$ star (Gaia EDR3 4080052973666546176\edit1{)} in the large photometric aperture used here). We searched the Gaia-EDR3 \citep{GaiaEDR3} catalog for additional contaminating field stars that were within a 6\arcsec\ radius. The only additional potential contaminants that had $G<17$ were a $G=16.6$ star with separation $2\farcs7$--$5\farcs4$ from the comet on 2021 October 21.38 and a $G=16.7$ star with separation $5\farcs3$--$5\farcs6$ from the comet on 2021 October 31.06. We can therefore be reasonably confident that a stellar contaminant is not the cause of any apparent brightening.

The timing of the second outburst is not well constrained. As noted above, our data on 2021 Oct 29 were contaminated by a bright star and was not usable. We searched the MPC database for additional data around this time but were unable to find any reports of total brightness before 2021 Nov 5. This lack of data for 2021 Oct 25 to 2021 Nov 01, when the first outburst was still declining and the second outburst occurred, makes the amplitude of the second outburst hard to judge. If we assume the LOOK data from 2021 Oct 31 at \rps$\sim13.9$ was during the rise, this put the outburst onset around 2021 Oct 30, giving an estimated $\sim0.9$\,mag increase in brightness from the prior measurement during this time. There was an independent announcement\footnote{\href{https://groups.io/g/comets-ml/message/30180}{comets-ml msg \#30180}} on 2021 Nov 06.72 of what is most likely this outburst by an amateur astronomer as a $\sim-0.4$\,mag outburst measured in a $5\farcs8$ aperture.

\begin{figure}
    \centering
    \includegraphics[width=\textwidth]{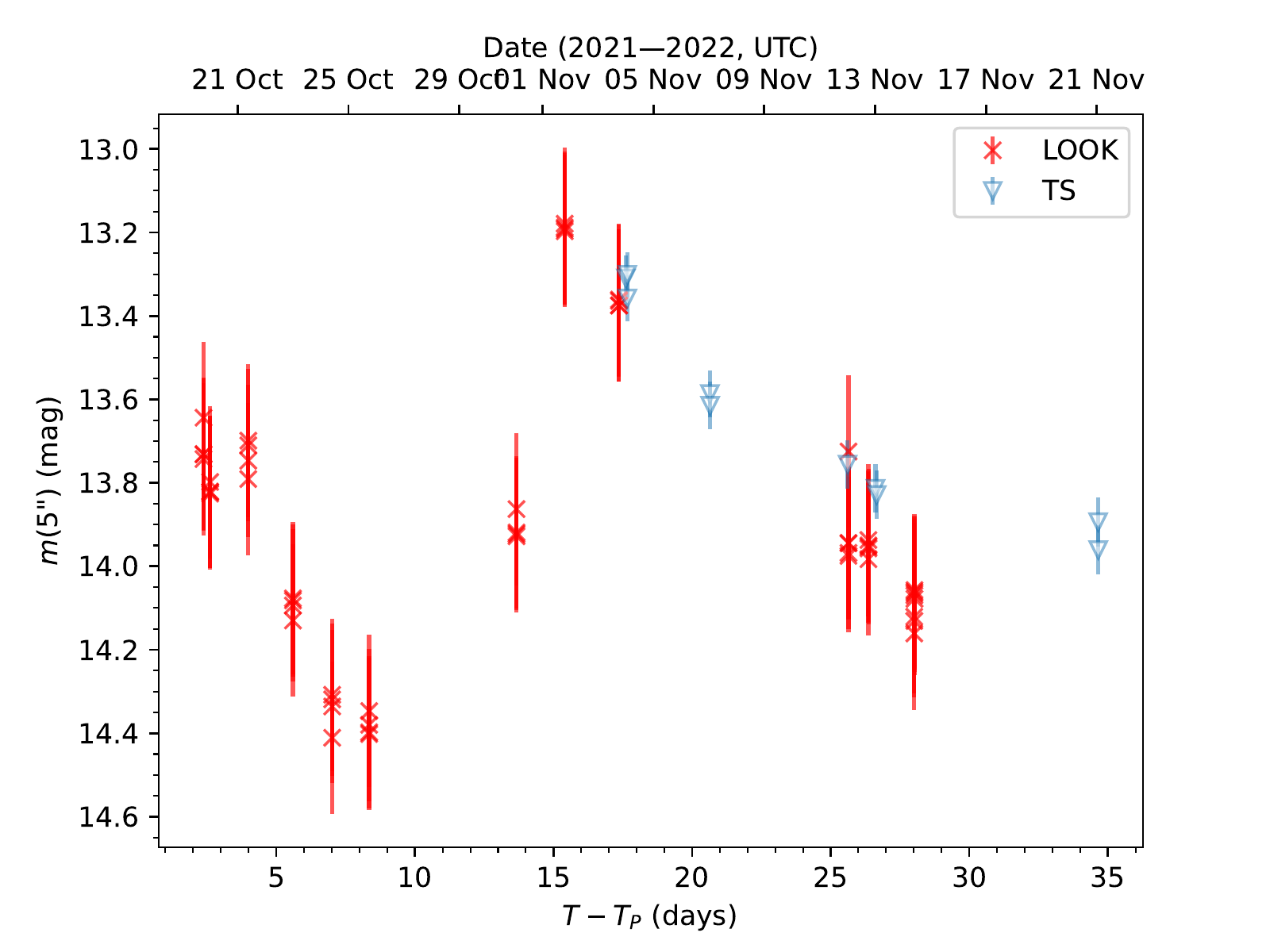}
    \caption{Lightcurve of comet 57P based on 5\arcsec{} radius \rps-band photometry from the LOOK Project and TRAPPIST South (TS) $R_c$-band photometry which has been offset by the color of the Sun: $R_c-\rps=-0.21$~mag \citep{Willmer2018}. Data from 2021 Oct. 29, which was affected by a bright stellar contaminant, has been removed.}
    \label{fig:57P}
\end{figure}

\begin{figure}
    \centering
    \includegraphics[width=\textwidth]{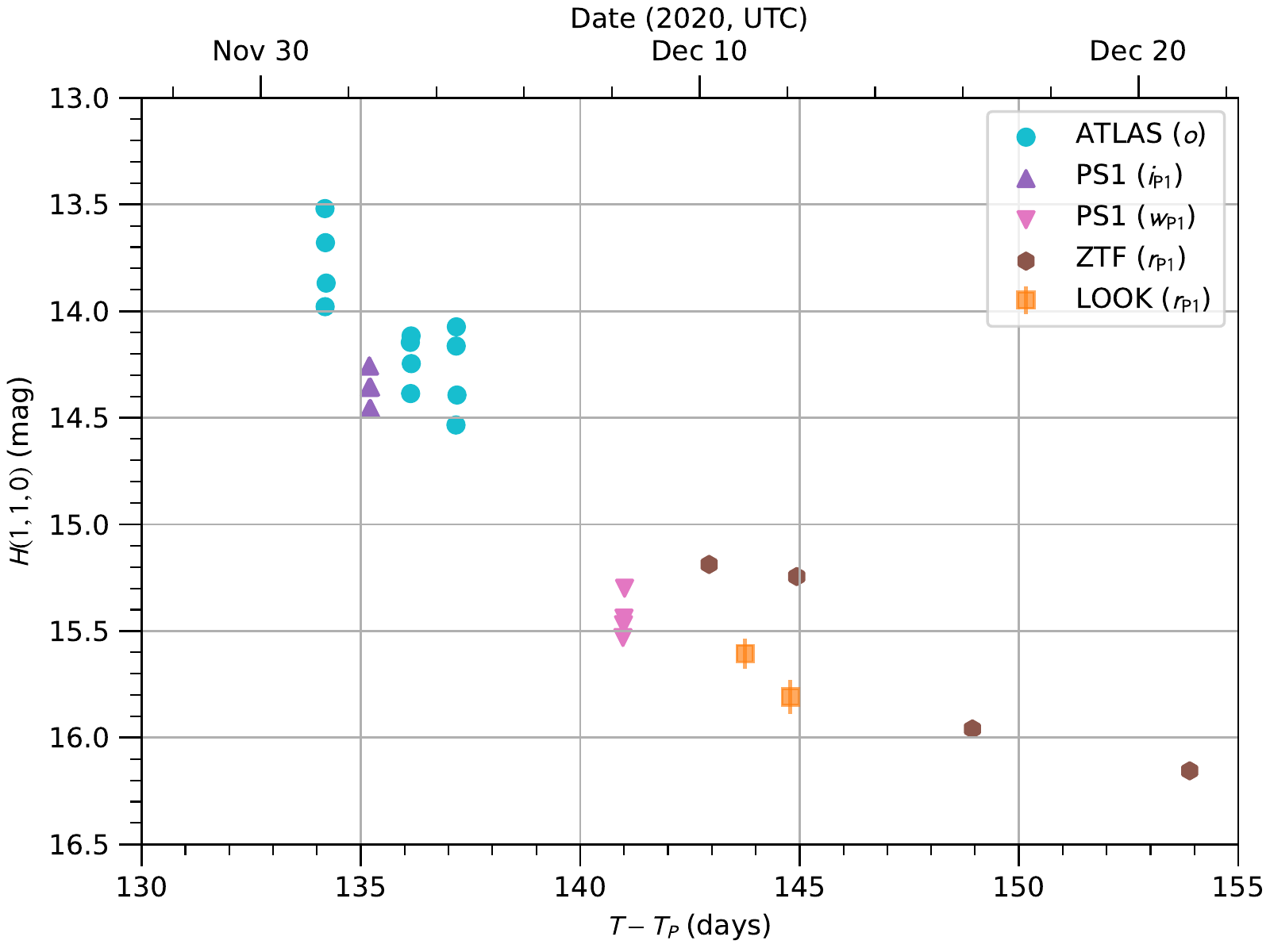}
    \caption{Absolute magnitude of comet P/2020 X1 (ATLAS) based on ATLAS \citep[$o$-band;][]{Tonry2018ATLAS}, PanSTARRS  \citep[PS1, $i_{\mathrm{P1}}$- and $w_{\mathrm{P1}}$-band;][]{Tonry2012PS1}, and ZTF \citep[$r_{\mathrm{P1}}$-band;][]{Bellm2019} data reported to the Minor Planet Center, and 5\arcsec{} radius \rps-band photometry from the LOOK Project.  No attempt has been made to correct for different aperture sizes and bandpasses.  However, the rapid decline in brightness, suggestive of an outburst, is clear.}
    \label{fig:p2020x1}
\end{figure}

\paragraph{P/2020 X1 (ATLAS)}
Comet P/2020 X1 is a JFC in a 9.6-year orbit with a perihelion distance of 2.87~au.  It was discovered during the course of the ATLAS sky survey. \citet{Fitzsimmons2020-P2020X1} report discovery, follow-up, and pre-discovery observations between 2020 Dec 01 and 08.  In the reported photometry, the comet's apparent magnitude is clearly increasing with time on day-long timescales.  LOOK follow-up observations comprise detections on 2020 Dec 11.035 and 12.081 at $\rps=19.67\pm0.07$ and $19.88\pm0.08$~mag, respectively (5\arcsec{} radius aperture).  In Fig.~\ref{fig:p2020x1}, we show ATLAS, PanSTARRS, ZTF, and LOOK project photometry of the comet as absolute magnitude, corrected using the Schleicher-Marcus phase function \citep{Schleicher2011}.  Neglecting bandpass and aperture size differences, the absolute magnitude increases $\sim$2~mag over an 11 day period, and more slowly thereafter, suggesting the comet was discovered during an outburst.

\subsubsection{Centaurs}
\protect\label{sec:centaurs}
Centaurs are recent escapees from the Kuiper belt and Trans-Neptunian region on giant planet crossing and scattering orbits. They serve as the likely progenitors of the Solar System's JFC population \citep{1997Sci...276.1670D,2020CeMDA.132...36D}. Centaurs have been observed to exhibit comet-like activity, which may be caused by drivers distinct from most short period comets \citep[e.g.][]{2009AJ....137.4296J,2012AJ....144...97G}. In 2021 August, the ATLAS survey identified a brightening in 95P/(2060) Chiron's apparent magnitude that was not consistent with rotational or orbital phase effects. This indicated a possible new cometary outburst or an enhanced phase of cometary activity in Chiron \citep{Dobson2021ATel, Dobson2021}. ATLAS observations found no definitive coma or extended features. As described by \citet{Dobson2021}, four 245s $w$-band follow-up exposures were taken on 2021 September 06 using the LCO 1-m telescope at Cerro Tololo Interamerican Observatory (CTIO). The telescope was tracked at Chiron's on-sky rate of motion, resulting in star trails of 0\farcs4 per exposure. The deeper LOOK observations were consistent with the ATLAS data, finding no extended PSF (point spread function) or visible coma/tail features. \edit1{It is unclear from the available LOOK data whether this was the start of a brand new outburst or an increase from a baseline activity held constant over the past 6 years. Deeper \edit2{exposures} and longer term data will be needed to try and resolve this situation once 95P becomes visible again from 2022 July.}

\subsection{Other observations}
\begin{figure}[ht]
    \centering
    \includegraphics[width=2.95in]{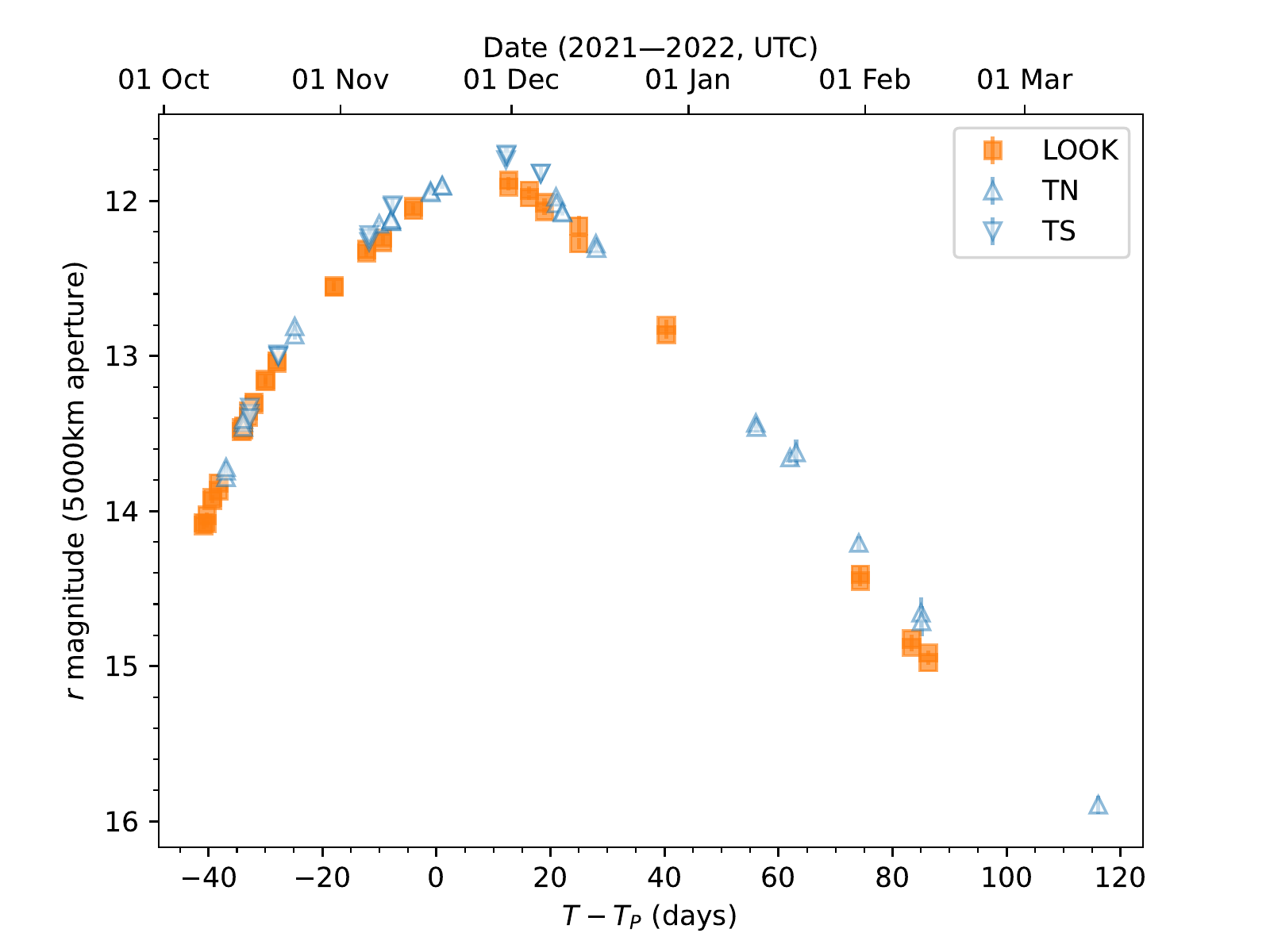}
    \includegraphics[width=2.95in]{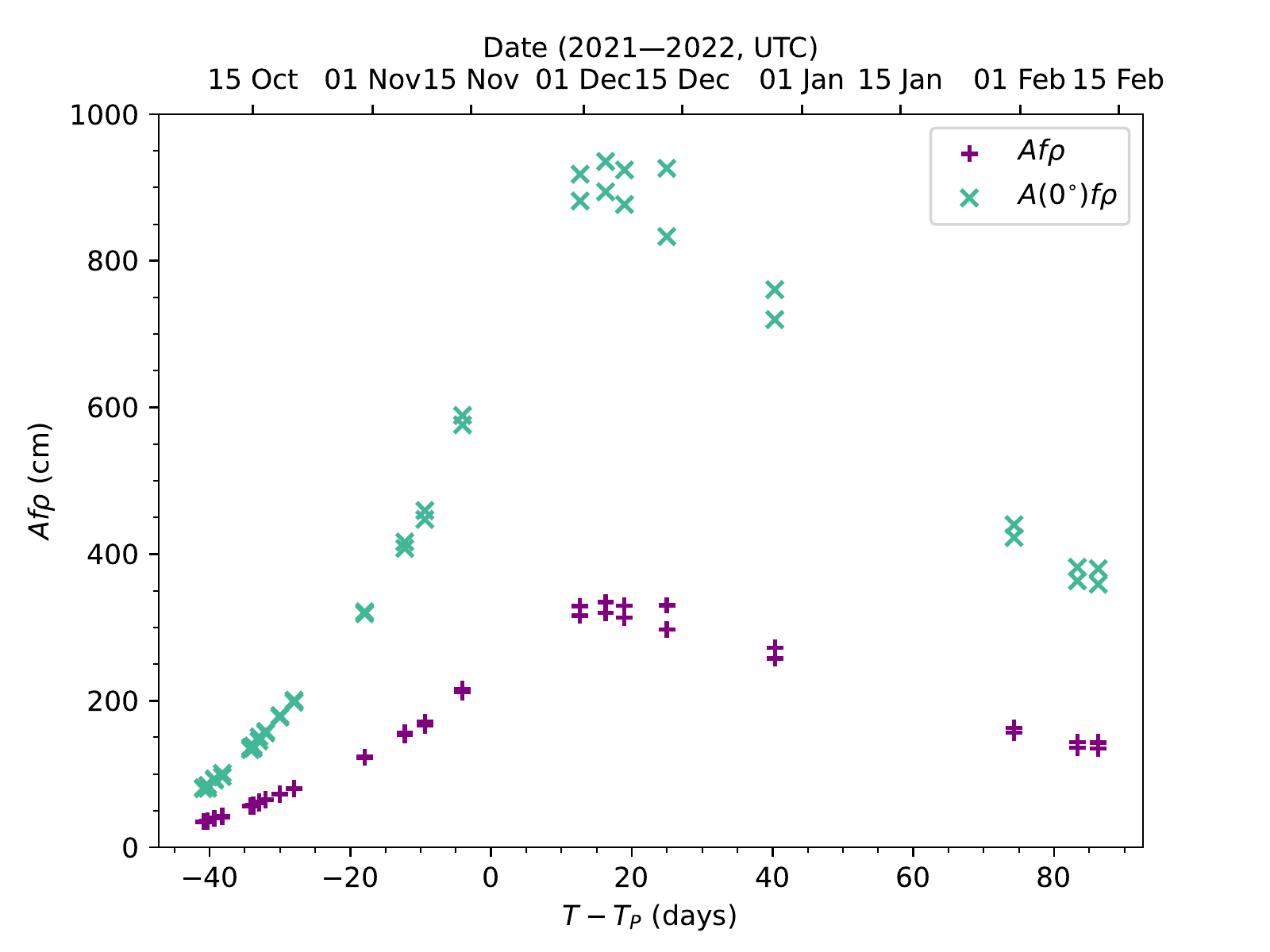}

    \caption{LOOK Project and TRAPPIST lightcurve (indicated by `TN' for TRAPPIST-North and `TS' for TRAPPIST-South) of comet 156P/Russell-LINEAR, measured within a 5,000\,km radius aperture at the comet distance.  LOOK \gps-band photometry has been offset by the measured color of the inner coma, $\gps-\rps=0.50\pm0.01$~mag, to produce effective \rps-band data. Also plotted (right figure) are the $Af\rho$ and $A(0^\circ)f\rho$ values from LOOK. Both plots show time as a function of time from perihelion ($T_p=24559171.313576862$ or 2020 Nov 17 19:31:33 TDB)}
    \label{fig:russell-linear}
\end{figure}

\paragraph{156P/Russell-LINEAR }
156P is a short period JFC which had a complicated discovery history \citep{McNaught2003-156P} having \edit1{been} discovered on four separate occasions with an apparently asteroidal appearance on three of those occasions. The comet has decreased in perihelion distance from $q\sim1.58$\,au in the 2014 perihelion passage to $q\sim1.33$\,au following a close pass ($\sim0.36$\,au) with Jupiter in 2018 March. This has also shortened the orbital period from 6.88 to 6.44 years. After a report on 2020 October 06 that the comet was found to be brighter than expected\footnote{\url{https://groups.io/g/comets-ml/message/29149}} and a follow-up note on the perturbation by Jupiter\footnote{\url{https://groups.io/g/comets-ml/message/29150}}, we initiated LOOK follow-up observations.  The subsequent data showed that the comet was not in outburst, but was undergoing enhanced activity.  We continued to observe this comet through 2021 February 12.  Photometry of the comet is presented in Fig.~\ref{fig:russell-linear}.  The LOOK data have been combined with data from the TRAPPIST-North and South telescopes \citep{Jehin2011} which have been transformed from $R$ to $r'$.
The combined lightcurve, along with $Af\rho$, shows a marked asymmetry in the lightcurve on either side of perihelion. This behavior could be due to the orientation of the active source regions on the surface of the comet toward Earth before and after perihelion. The gas activity of the comet shows \edit1{the same asymmetric shape as the dust activity} on both sides of perihelion (Moulane et al. private communication). This kind of asymmetry in the comet's activity has been seen in many LPCs and JFCs. Narrow-band photometry of comet 156P shows clear detection of radicals such OH, CN, C$_2$ and C$_3$ in the optical, indicating that the comet has a typical composition given the gas production rates ratios \citep{Jehin2020ATel14101}.

\begin{figure}[ht]
 \centering \includegraphics[scale=0.30]{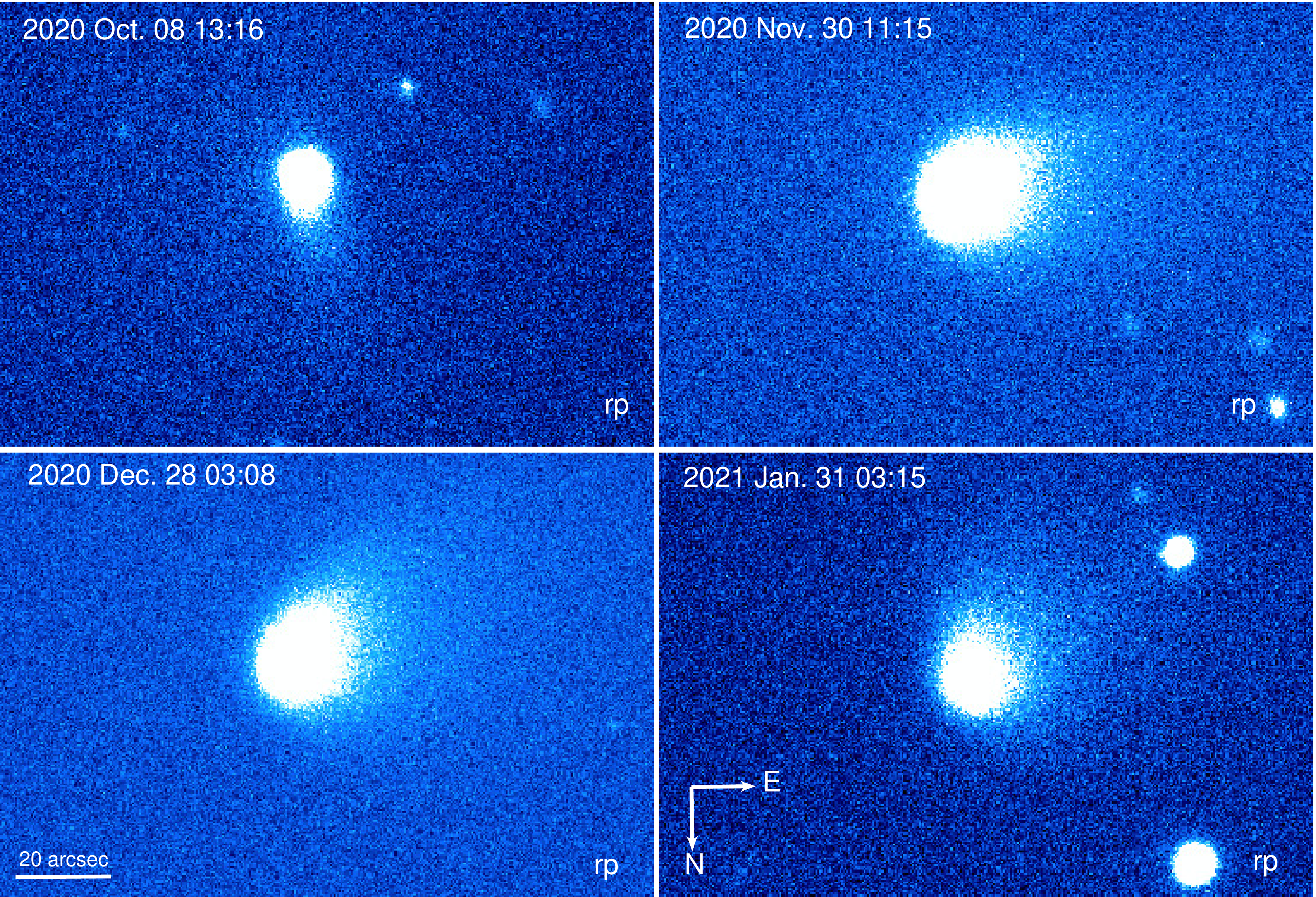}
 \caption{Evolution of 156P coma morphology before and after perihelion in $r'$ images from LOOK. The orientation and scale are given at the bottom of the lower images.}
 \label{156P-coma}
 \end{figure}

Figure \ref{156P-coma} shows the $r'$ coma evolution with time. \edit1{The shape of the coma was asymmetric and it has changed} throughout the monitoring window. 156P reached its peak activity on 2020 Nov 30 (12 days after perihelion), when the coma was the brightest. There are possible hints of substructure in the coma in the 2020 Nov and Dec images. Since then, the comet faded gradually.

We performed a search \edit2{of} the magnitudes reported to the Minor Planet Center and the Comet OBServations (COBS\footnote{\url{https://www.cobs.si/}}) database for data from previous perihelion passages. This resulted in 38 observations from 2007 Sep 11 -- 2009 Apr 19 from the 2007 perihelion and 26 observations from 2013 July 05 -- 2015 April 16 from the 2014 perihelion. These are dwarfed in number by the almost 1900 observations (so far) from the 2020 perihelion.

\begin{figure}
    \centering
    \includegraphics[width=5in]{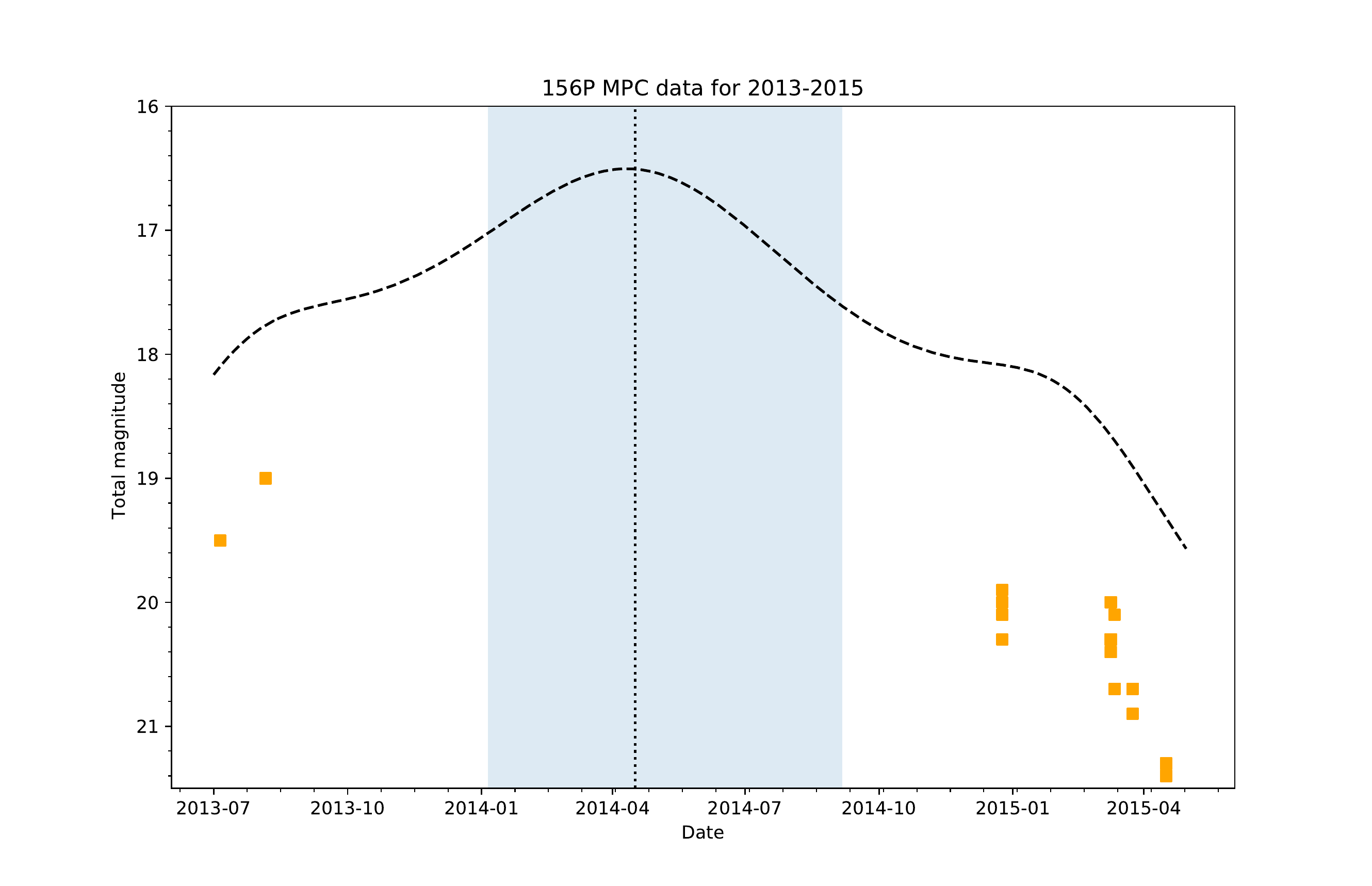}
    \caption{Total magnitudes for 156P from the Minor Planet Center database (orange squares) for the 2013--2015 apparition along with the predicted total magnitude from JPL Horizons (black dashed line). The time of perihelion is indicated by the vertical dotted line and the time when the solar elongation was $<40^\circ$ is shown by the blue shaded area.}
    \label{fig:russell-linear-prevapp}
\end{figure}

The data from the 2013--2015 apparition, along with the predicted evolution of the total magnitude are plotted in Figure~\ref{fig:russell-linear-prevapp}. The JPL Horizons predicted total magnitude uses the IAU model for comet brightness:
\begin{displaymath}
\textrm{Total mag.}= M_1 + 5\log_{10}(\Delta) + k_1\log_{10}(r)
\end{displaymath}
with $M_1=12.7, k_1=8.5$, where $\Delta$ and $r$ are the Earth-comet and Sun-comet distances respectively. An independent fit to all the 1986--2015 data (prior to the current brighter apparition and the encounter with Jupiter) produced a model with $M_1=14.3, k_1=10.0$, which significantly under-predicted the observed brightness during 2013--2015. The data are very sparse and consist mainly of data from the NEO surveys but appear to indicate that the intrinsic brightness of the comet was significantly ($\sim1.25$\,mag) fainter than expected based on the JPL fit to the 1993--2022 data alone (which is dominated by the large number of post-2020 observations). This may have contributed to the dearth of observations reported over the approximately 16 months around perihelion, along with a long period of unobservability when the comet was close to the Sun.

We speculate that the increased activity of 156P in 2020 may have been initiated by the orbit change since the previous perihelion passage in 2014 April. Some possible mechanisms for the increased activity include increased insolation and resulting deeper propagation of the thermal wave or a shift in the pole orientation that has resulted in changing seasonal illumination of the source regions. Another possibility is that a \edit1{torque was applied to the nucleus} due to the close approach to Jupiter that rearranged/exposed previously buried volatiles, although this seems somewhat unlikely given the nominal close approach distance of 0.36\,au.

\paragraph{C/2021 A1 (Leonard)}
Oort cloud comet C/2021 A1 (Leonard) reached perihelion on 2022 January 3 at $q=0.615$~au. Although Leonard is not a DNC, we included it in our sample since its heliocentric distance changed by nearly a factor of 6 between our earliest observation on 2021 May 25 and perihelion, making it a good target for comparison with DNC brightness evolution. Its generally small solar elongation near perihelion prevented extensive observations from any one facility, but a brief window before perigee (0.23~au on 2021 December 12) allowed regular monitoring with LCO, and we obtained 13 epochs of $g'$ and $r'$ imaging from 2021 November 6 to December 6.

\begin{figure}
    \centering
    \includegraphics[width=6in]{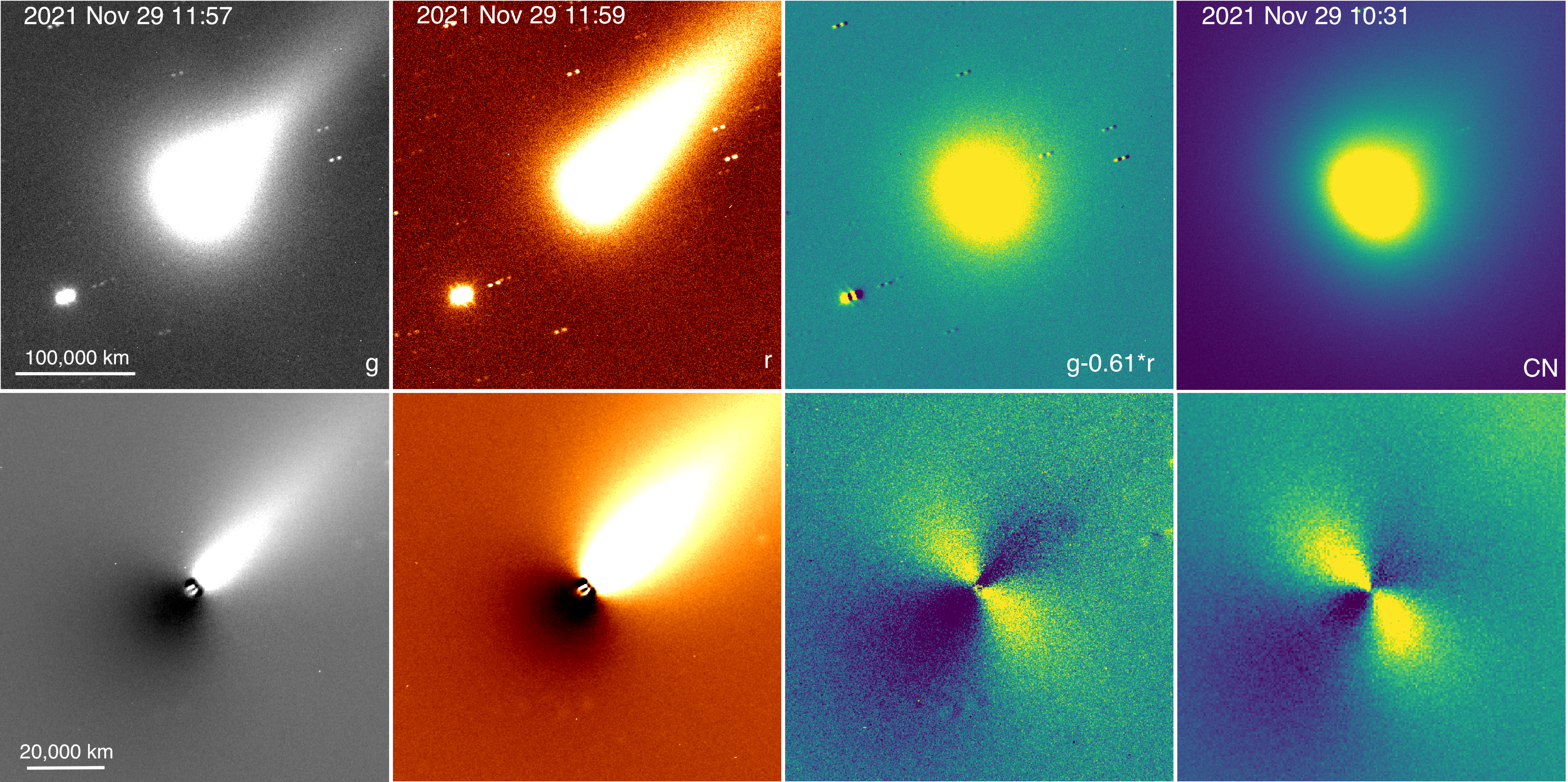}
    \caption{
    Unenhanced (top row) and enhanced (bottom row) images of C/2021 A1 (Leonard) in various filters on 2021 November 29. All images are centered on the opto-center of the comet and have north up and east to the left. All images on the top row are 325,000 km across while all images on the bottom row are 100,000 km across and have been enhanced by subtraction of an azimuthal median profile. Light is bright and dark is faint in all panels.
    The \edit1{first and second panels in each row are the $g'$ and $r'$ images respectively, the} third panel in each row is a synthetic ``gas'' image created by scaling an $r'$ image (median of all images taken that night) by the estimated dust color and subtracting this scaled image from a median $g'$ image. As discussed in the text, the morphology of this synthetic gas image closely matches that of the
    CN gas image (\edit1{fourth column;} acquired by B.~Skiff using the Lowell Observatory 42in telescope) and is very different from the ``dust'' ($r'$-band) image.
    Two jets can be seen oriented near positional angles 40$^\circ$ and 220$^\circ$ in the enhanced gas images, while the tail is to the northwest in the dust images. A very faint residual dust tail is still seen near the northwest corner in the gas images.
}
    \label{fig:2021A1_cn}
\end{figure}

C/2021 A1 displayed distinctly different morphology between $g'$ and $r'$ when it was brightest in late 2021, prior to perihelion on 2022 January 03. Since the $g'$ filter contains bright C$_2$ emission bands and fainter C$_3$ emission bands (see Figure~\ref{fig:comet_spectrum}), while $r'$ is free of bright emission bands, we constructed a synthetic ``gas'' image by scaling an $r'$ image and subtracting it from the contemporaneously obtained $g'$ image. The scaling was conducted by eye, with the scaling coefficient determined by looking for the disappearance of the dust tail in the ``gas'' image.  The best determined value was 0.61, but values within $\pm0.02$ produced indistinguishable results\edit1{. The scale images were in flux units calibrated via the AB magnitude system (i.e., $F_{\nu}(g') - 0.61{\times}F_{\nu}(r')$); converting from flux to magnitudes implies that the dust color was} $\gps-\rps$=0.54$\pm$0.05~mag, including calibration uncertainties.  Smaller scaling coefficients resulted in a visible excess of positive ($g'$) tail, while larger scaling coefficients resulted in an excess of negative ($r'$) tail; our final scaling coefficient choice varied from night to night, but the resulting morphology was essentially unchanged for values within 0.02~mag.  In Fig.~\ref{fig:2021A1_cn}, we show example $g'$ and $r'$ images from 2021 November 29 ($r_h=0.95$~au, $\Delta=0.58$~au).  The LCO telescope used was not optimally focused at the time of the data acquisition, but the broad structures (coma and tail) are not adversely affected.  The dust color, $\gps-\rps$=0.54~mag, is considerably redder than the total coma color measured within a 5000~km (10\arcsec) radius aperture: $\gps-\rps=0.21\pm0.04$~mag. This implies that the gas contamination is $\sim$26\% in the $g'$ filter for this observation and aperture size.  TRAPPIST narrow-band observations on 2021 December 19 show that the comet has dust to gas ratio typical of most comets, with an OH to dust $Af\rho$  production rate ratio of $Q$(OH)/$Af\rho$=  9$\times$10$^{25}$~molecules~s$^{-1}$~cm$^{-1}$ \citep{Jehin2021ATel15128}.

Figure~\ref{fig:2021A1_cn} also shows an image taken 1.5 hr earlier using the narrowband CN filter from the HB comet filter set \citep{Farnham2000} with Lowell Observatory's 42-in telescope and provided by B.~Skiff. The gas morphology can be compared with that seen in our best synthetic ``gas'' image by looking at the last two panels of the bottom row. Both images are 100,000 km across, centered on the opto-center, and have been enhanced by subtraction of an azimuthal median in order to remove the bulk brightness and emphasize subtle brightness variations. This reveals two apparent gas jets of comparable brightness and extent, one to the northeast and the other to the southwest. Although the CN image shown has not been properly decontaminated of dust, previous work \citep{Knight2011} has found that the dust contamination is negligible for comets with ``typical'' gas-to-dust ratios like Leonard. Furthermore, CN and C$_2$ coma morphology tend to be very similar \citep[e.g.,][]{Knight2013,Knight2021}, so a first order comparison is reasonable to assess the technique. The figure also shows our $g'$ and $r'$ images at the same scale and enhanced by the same technique. The dust morphology is completely different from the gas, being dominated by the tail towards the northwest.

The orientation and extent of the jets is similar in both images, giving us confidence that the technique is useful for a qualitative assessment of gas in sufficiently bright comets. Small variations within a few thousand km of the center are art\edit1{i}facts of the processing and image quality and should be ignored. There are subtle variations in the jet morphology, with the CN jets tilting slightly towards the southeast (in the sunward direction), while our ``gas'' jets show a slight tilt towards the northwest (in the tailward direction). It is possible that these differences are real and indicative of rotational variations during the intervening 1.5 hr, but they might also be due to the different gas species involved or to the processing technique. The ability of this technique to  monitor rotational variation of gas morphology will be explored in a future paper; for now we simply note that it appears viable.

\section{Discussion and Conclusions}
\label{sec:summary}

We have described the scientific goals and motivation for the \LOOKfull, a three year, long-term status proposal on the \LCONet, along with details on our target selection, monitoring, outburst response, and data reduction steps. We have also outlined some of the initial results from the first $\sim$16 months of operation.

Completion of observing (at the end of 2023 July) and comprehensive analysis of the full three year dataset will be needed before we can make definitive statements on the behavior of DNCs, specifically with regards to changes in activity with distance from the Sun. These results will be valuable to feed into the planning and target selection for ESA's Comet Interceptor mission to visit an, ideally, dynamically new comet after launch in 2029. \edit1{A more complete understanding of the evolution of the activity behavior of DNCs as a function of distance and target properties could aid the determination of the best target for Comet Interceptor should we find ourselves in the fortunate situation of having a choice of targets that are reachable by the spacecraft.} \edit1{Raw and BCD images and data products from the LOOK observing program, as with all data taken on the \LCONet, are available from the public LCO Science Archive as detailed in the acknowledgements. Additional data such as spectra from other facilities and catalogs of derived properties such as photometry will be made available from the LOOK project website following publication of papers containing the data or the end of observing program.}

The recent discovery of two confirmed interstellar objects (1I/`Oumuamua and 2I/Borisov) and the very large diameter \BB along with its activity even at large heliocentric distances, shows the potential of what can be discovered through the combination of large area sky surveys coupled with rapid data reduction, analysis, and alert generation. This potential is driving a lot of the development of survey and analysis strategy for the Vera C.\ Rubin Observatory and how its Legacy Survey of Space and Time (LSST) will be carried out. In addition, it is driving thinking and evolution of the transient follow-up ecosystem for existing and future surveys such as LSST, including developments such as the Astrophysical Events Observatories Network (AEON; \citealt{Street2020AEON}). The first iteration of AEON, with the use of the SOAR 4.1\,m telescope in Chile being scheduled in a remote queue mode along with the rest of the \LCONet by the LCO scheduler, has already been successfully utilized by LOOK project members to study fainter DNCs \citep[preliminary results were presented by ][]{Holt2021}. As shown in Section~\ref{sec:mtjohn}, there is also great potential for integrating classically scheduled and operated telescopes into a follow-up ecosystem to provide additional observational resources for follow-up. Coordinated observations and management of large observing programs utilizing robotic (LCO), partially queue-based (SOAR), and non-robotic telescopes (Mt.\ John) is possible through a TOM system as discussed in Section~\ref{sec:obssched}.

The LOOK Project data set so far, presented here, demonstrates that small transient events may be readily discovered and studied in cometary comae with the \LCONet.  As discussed in Section~\ref{sec:outburst-comets}, \edit2{28} outbursts \edit2{of 14} comets have been discovered or studied in the first $\sim16$ months of the \LOOK. In addition to these outbursts of the more heavily monitored and larger sample of bright comets, we have also presented evidence of quiescent activity on an active asteroid (Section~\ref{sec:outburst-ast}) and outbursts on a Centaur (Section~\ref{sec:centaurs}). These outburst detections are primarily coming from ZTF, which covers the northern sky in two bands every two nights, supplemented by outburst discoveries from LOOK data and personnel analysis as well as amateur observers. Although there is a generalized alert system for ZTF \citep{Patterson2019} which should include active asteroids and comets, in the desire to avoid sending alerts for image artifacts and the bias towards stellar science within ZTF, this means that extended objects such as comet or activated asteroids are filtered out of the alert stream. This reinforces the need for the next generation of alert brokers \citep[e.g.][]{Narayan2018, Smith2019Lasair, Forster2021ALeRCE} for Rubin Observatory to be able \edit1{to} handle all science cases, including moving and extended objects.   There is the strong possibility of additional alerts for the southern sky to come from the ATLAS 3 \& 4 telescopes in Chile and South Africa with their entrance into regular operations during 2022. Although there is not a general alert system for ATLAS, we can expect an increased number of outburst discoveries in the southern hemisphere where LCO has more telescopes and a more even longitudinal spread of telescopes, allowing a more rapid response and continuous coverage of outbursts.

Although COVID-19 pandemic-related delays have pushed the start of survey operations and alert generation from the Rubin Observatory beyond the end date of the \LOOK, the discovery of new exceptional objects (such as C/2014 UN$_{271}$ Bernardinelli-Bernstein) and the alerts from the already operating surveys (or those that will start soon) will provide valuable targets and input for developing, operating, and testing the broader transient follow-up infrastructure for planetary science in preparation for the era of a larger volume of objects and alerts once Rubin observations start. The determination of the rate, behavior, and evolution of outbursts on the variety of Solar System objects covered by LOOK, along with characterization of the onset and evolution of activity on DNCs will form a valuable resource for the future.

\begin{acknowledgements}

We appreciate the work and dedication of the amateur astronomical community and their contributions to the study of transient phenomena in small solar system objects. We thank Brian Skiff for providing the CN image of C/2021 A1 Leonard.

MSPK and SP acknowledge support from the NASA Solar System Observations program (80NSSC20K0673). MTB appreciates support by the Rutherford Discovery Fellowships from New Zealand Government funding, administered by the Royal Society Te Ap\={a}rangi. MES, MMD, and CS were supported in part by the UK Science Technology Facilities Council (STFC) grants ST/V000691/1, ST/V506990/1, and ST/V000586/1, respectively.
MMK was supported by NASA Solar System Observations program grant 80HQTR20T0060.
Development of NEOexchange was funded in part by the NASA Near Earth Objects Observations program through  grants NNX14AM98G and 80NSSC18K0848 to \LCOfull is acknowledged.

This work makes use of observations from the Las Cumbres Observatory global telescope network.  Observations with the LCOGT 1m were obtained as part of the LCO Outbursting Objects Key (LOOK) Project (KEY2020B-009). Some observations in this paper are based on observations made with the MuSCAT3 instrument, developed by the Astrobiology Center and under financial supports by JSPS KAKENHI (JP18H05439) and JST PRESTO (JPMJPR1775), at Faulkes Telescope North on Maui, HI, operated by the Las Cumbres Observatory. TRAPPIST-South is a project funded by the Belgian Fonds (National) de la Recherche Scientifique (F.R.S.-FNRS) under grant PDR T.0120.21. TRAPPIST-North is a project funded by the University of Li\`{e}ge, in collaboration with the Cadi Ayyad University of Marrakech (Morocco). E. Jehin is FNRS Senior Research Associate.

This research has made use of data and services provided by the International Astronomical Union's Minor Planet Center.

This research has made use of NASA's Astrophysics Data System Bibliographic Services.
Data Access: Raw Data Products (and Basic Calibrated Data products as described in Section~\ref{sec:dataproc}) supporting this study are available from the LCO Science Archive at \url{https://archive.lco.global} using the KEY2020B-009 proposal code following a 12 month embargo/proprietary period.

\end{acknowledgements}

\facilities{LCOGT, FTN, MtJohn:1.8m, TRAPPIST}

\software{astropy \citep{astropy18}, sbpy \citep{Mommert2019}, astroquery \citep{astroquery2019}, JPL Horizons \citep{Giorgini1996}, SEP \citep{Barbary2016-SEP}, DS9 \citep{Joye2003-DS9}, reproject \citep{Robitaille2018-reproject}, photutils \citep{Bradley2021-photutils1.1.0}, calviacat \citep{Kelley2021-calviacat}, ccdproc \citep{Craig2021-ccdproc2.2.0}, BANZAI \citep{McCully2018BANZAI}}


\bibliography{look_overview,lco_solsys}{}

\begin{thebibliography}{}
\expandafter\ifx\csname natexlab\endcsname\relax\def\natexlab#1{#1}\fi
\providecommand{\url}[1]{\href{#1}{#1}}
\providecommand{\dodoi}[1]{doi:~\href{http://doi.org/#1}{\nolinkurl{#1}}}
\providecommand{\doeprint}[1]{\href{http://ascl.net/#1}{\nolinkurl{http://ascl.net/#1}}}
\providecommand{\doarXiv}[1]{\href{https://arxiv.org/abs/#1}{\nolinkurl{https://arxiv.org/abs/#1}}}

\bibitem[{{A'Hearn} {et~al.}(1995){A'Hearn}, {Millis}, {Schleicher}, {Osip}, \&
  {Birch}}]{A'Hearn1995}
{A'Hearn}, M.~F., {Millis}, R.~C., {Schleicher}, D.~O., {Osip}, D.~J., \&
  {Birch}, P.~V. 1995, \icarus, 118, 223, \dodoi{10.1006/icar.1995.1190}

\bibitem[{{A'Hearn} {et~al.}(1984){A'Hearn}, {Schleicher}, {Millis}, {Feldman},
  \& {Thompson}}]{A'Hearn1984}
{A'Hearn}, M.~F., {Schleicher}, D.~G., {Millis}, R.~L., {Feldman}, P.~D., \&
  {Thompson}, D.~T. 1984, \aj, 89, 579, \dodoi{10.1086/113552}

\bibitem[{{A'Hearn} {et~al.}(2011){A'Hearn}, {Belton}, {Delamere}, {Feaga},
  {Hampton}, {Kissel}, {Klaasen}, {McFadden}, {Meech}, {Melosh}, {Schultz},
  {Sunshine}, {Thomas}, {Veverka}, {Wellnitz}, {Yeomans}, {Besse}, {Bodewits},
  {Bowling}, {Carcich}, {Collins}, {Farnham}, {Groussin}, {Hermalyn}, {Kelley},
  {Kelley}, {Li}, {Lindler}, {Lisse}, {McLaughlin}, {Merlin}, {Protopapa},
  {Richardson}, \& {Williams}}]{A'Hearn2011}
{A'Hearn}, M.~F., {Belton}, M. J.~S., {Delamere}, W.~A., {et~al.} 2011,
  Science, 332, 1396, \dodoi{10.1126/science.1204054}

\bibitem[{{Astropy Collaboration} {et~al.}(2018){Astropy Collaboration},
  {Price-Whelan}, {Sip{\H{o}}cz}, {G{\"u}nther}, {Lim}, {Crawford}, {Conseil},
  {Shupe}, {Craig}, {Dencheva}, {Ginsburg}, {VanderPlas}, {Bradley},
  {P{\'e}rez-Su{\'a}rez}, {de Val-Borro}, {Aldcroft}, {Cruz}, {Robitaille},
  {Tollerud}, {Ardelean}, {Babej}, {Bach}, {Bachetti}, {Bakanov}, {Bamford},
  {Barentsen}, {Barmby}, {Baumbach}, {Berry}, {Biscani}, {Boquien}, {Bostroem},
  {Bouma}, {Brammer}, {Bray}, {Breytenbach}, {Buddelmeijer}, {Burke},
  {Calderone}, {Cano Rodr{\'\i}guez}, {Cara}, {Cardoso}, {Cheedella}, {Copin},
  {Corrales}, {Crichton}, {D'Avella}, {Deil}, {Depagne}, {Dietrich}, {Donath},
  {Droettboom}, {Earl}, {Erben}, {Fabbro}, {Ferreira}, {Finethy}, {Fox},
  {Garrison}, {Gibbons}, {Goldstein}, {Gommers}, {Greco}, {Greenfield},
  {Groener}, {Grollier}, {Hagen}, {Hirst}, {Homeier}, {Horton}, {Hosseinzadeh},
  {Hu}, {Hunkeler}, {Ivezi{\'c}}, {Jain}, {Jenness}, {Kanarek}, {Kendrew},
  {Kern}, {Kerzendorf}, {Khvalko}, {King}, {Kirkby}, {Kulkarni}, {Kumar},
  {Lee}, {Lenz}, {Littlefair}, {Ma}, {Macleod}, {Mastropietro}, {McCully},
  {Montagnac}, {Morris}, {Mueller}, {Mumford}, {Muna}, {Murphy}, {Nelson},
  {Nguyen}, {Ninan}, {N{\"o}the}, {Ogaz}, {Oh}, {Parejko}, {Parley}, {Pascual},
  {Patil}, {Patil}, {Plunkett}, {Prochaska}, {Rastogi}, {Reddy Janga},
  {Sabater}, {Sakurikar}, {Seifert}, {Sherbert}, {Sherwood-Taylor}, {Shih},
  {Sick}, {Silbiger}, {Singanamalla}, {Singer}, {Sladen}, {Sooley},
  {Sornarajah}, {Streicher}, {Teuben}, {Thomas}, {Tremblay}, {Turner},
  {Terr{\'o}n}, {van Kerkwijk}, {de la Vega}, {Watkins}, {Weaver}, {Whitmore},
  {Woillez}, {Zabalza}, \& {Astropy Contributors}}]{astropy18}
{Astropy Collaboration}, {Price-Whelan}, A.~M., {Sip{\H{o}}cz}, B.~M., {et~al.}
  2018, \aj, 156, 123, \dodoi{10.3847/1538-3881/aabc4f}

\bibitem[{Barbary(2016)}]{Barbary2016-SEP}
Barbary, K. 2016, Journal of Open Source Software, 1, 58,
  \dodoi{10.21105/joss.00058}

\bibitem[{{Bellm} {et~al.}(2019){Bellm}, {Kulkarni}, {Graham}, {Dekany},
  {Smith}, {Riddle}, {Masci}, {Helou}, {Prince}, {Adams}, {Barbarino},
  {Barlow}, {Bauer}, {Beck}, {Belicki}, {Biswas}, {Blagorodnova}, {Bodewits},
  {Bolin}, {Brinnel}, {Brooke}, {Bue}, {Bulla}, {Burruss}, {Cenko}, {Chang},
  {Connolly}, {Coughlin}, {Cromer}, {Cunningham}, {De}, {Delacroix}, {Desai},
  {Duev}, {Eadie}, {Farnham}, {Feeney}, {Feindt}, {Flynn}, {Franckowiak},
  {Frederick}, {Fremling}, {Gal-Yam}, {Gezari}, {Giomi}, {Goldstein},
  {Golkhou}, {Goobar}, {Groom}, {Hacopians}, {Hale}, {Henning}, {Ho}, {Hover},
  {Howell}, {Hung}, {Huppenkothen}, {Imel}, {Ip}, {Ivezi{\'c}}, {Jackson},
  {Jones}, {Juric}, {Kasliwal}, {Kaspi}, {Kaye}, {Kelley}, {Kowalski},
  {Kramer}, {Kupfer}, {Landry}, {Laher}, {Lee}, {Lin}, {Lin}, {Lunnan},
  {Giomi}, {Mahabal}, {Mao}, {Miller}, {Monkewitz}, {Murphy}, {Ngeow},
  {Nordin}, {Nugent}, {Ofek}, {Patterson}, {Penprase}, {Porter}, {Rauch},
  {Rebbapragada}, {Reiley}, {Rigault}, {Rodriguez}, {van Roestel}, {Rusholme},
  {van Santen}, {Schulze}, {Shupe}, {Singer}, {Soumagnac}, {Stein}, {Surace},
  {Sollerman}, {Szkody}, {Taddia}, {Terek}, {Van Sistine}, {van Velzen},
  {Vestrand}, {Walters}, {Ward}, {Ye}, {Yu}, {Yan}, \& {Zolkower}}]{Bellm2019}
{Bellm}, E.~C., {Kulkarni}, S.~R., {Graham}, M.~J., {et~al.} 2019, \pasp, 131,
  018002, \dodoi{10.1088/1538-3873/aaecbe}

\bibitem[{Bernardinelli \& Bernstein(2021)}]{Bernardinelli2021MPEC}
Bernardinelli, P., \& Bernstein, G. 2021, MPEC M53.
\newblock \url{https://minorplanetcenter.net/mpec/K21/K21M53.html}

\bibitem[{{Bernardinelli} {et~al.}(2021){Bernardinelli}, {Bernstein}, {Montet},
  {Weryk}, {Wainscoat}, {Aguena}, {Allam}, {Andrade-Oliveira}, {Annis},
  {Avila}, {Bertin}, {Brooks}, {Burke}, {Carnero Rosell}, {Carrasco Kind},
  {Carretero}, {Cawthon}, {Conselice}, {Costanzi}, {da Costa}, {Pereira}, {De
  Vicente}, {Diehl}, {Everett}, {Ferrero}, {Flaugher}, {Frieman},
  {Garc{\'\i}a-Bellido}, {Gaztanaga}, {Gerdes}, {Gruen}, {Gruendl}, {Gschwend},
  {Gutierrez}, {Hinton}, {Hollowood}, {Honscheid}, {James}, {Kuehn},
  {Kuropatkin}, {Lahav}, {Maia}, {Marshall}, {Menanteau}, {Miquel}, {Morgan},
  {Ogando}, {Paz-Chinch{\'o}n}, {Pieres}, {Malag{\'o}n}, {Rodriguez-Monroy},
  {Romer}, {Roodman}, {Sanchez}, {Schubnell}, {Serrano}, {Sevilla-Noarbe},
  {Smith}, {Soares-Santos}, {Suchyta}, {Swanson}, {Tarle}, {To}, {Troxel},
  {Varga}, {Walker}, {Zhang}, \& {DES Collaboration}}]{Bernardinelli2021}
{Bernardinelli}, P.~H., {Bernstein}, G.~M., {Montet}, B.~T., {et~al.} 2021,
  \apjl, 921, L37, \dodoi{10.3847/2041-8213/ac32d3}

\bibitem[{{Bodewits} {et~al.}(2011){Bodewits}, {Kelley}, {Li}, {Landsman},
  {Besse}, \& {A'Hearn}}]{Bodewits2011}
{Bodewits}, D., {Kelley}, M.~S., {Li}, J.~Y., {et~al.} 2011, \apjl, 733, L3,
  \dodoi{10.1088/2041-8205/733/1/L3}

\bibitem[{Bradley {et~al.}(2021)Bradley, Sip{\H o}cz, Robitaille, Tollerud,
  Vin{\'{\i}}cius, Deil, Barbary, Wilson, Busko, G{\"u}nther, Cara, Conseil,
  Bostroem, Droettboom, Bray, Bratholm, Lim, Barentsen, Craig, Pascual, Perren,
  Greco, Donath, de~Val-Borro, Kerzendorf, Bach, Weaver, D'Eugenio, Souchereau,
  \& Ferreira}]{Bradley2021-photutils1.1.0}
Bradley, L., Sip{\H o}cz, B., Robitaille, T., {et~al.} 2021, astropy/photutils:
  1.1.0, 1.1.0,  Zenodo, \dodoi{10.5281/zenodo.4624996}

\bibitem[{{Brown} {et~al.}(2013){Brown}, {Baliber}, {Bianco}, {Bowman},
  {Burleson}, {Conway}, {Crellin}, {Depagne}, {De Vera}, {Dilday}, {Dragomir},
  {Dubberley}, {Eastman}, {Elphick}, {Falarski}, {Foale}, {Ford}, {Fulton},
  {Garza}, {Gomez}, {Graham}, {Greene}, {Haldeman}, {Hawkins}, {Haworth},
  {Haynes}, {Hidas}, {Hjelstrom}, {Howell}, {Hygelund}, {Lister}, {Lobdill},
  {Martinez}, {Mullins}, {Norbury}, {Parrent}, {Paulson}, {Petry}, {Pickles},
  {Posner}, {Rosing}, {Ross}, {Sand}, {Saunders}, {Shobbrook}, {Shporer},
  {Street}, {Thomas}, {Tsapras}, {Tufts}, {Valenti}, {Vander Horst}, {Walker},
  {White}, \& {Willis}}]{Brown2013LCOGT}
{Brown}, T.~M., {Baliber}, N., {Bianco}, F.~B., {et~al.} 2013, \pasp, 125,
  1031, \dodoi{10.1086/673168}

\bibitem[{Buzzi {et~al.}(2021)Buzzi, Demetz, Aletti, Veres, Lister, Weryk,
  Nakano, \& Green}]{Buzzi2021}
Buzzi, L., Demetz, L., Aletti, A., {et~al.} 2021, CBET, 4989

\bibitem[{{Chambers} {et~al.}(2016){Chambers}, {Magnier}, {Metcalfe},
  {Flewelling}, {Huber}, {Waters}, {Denneau}, {Draper}, {Farrow}, {Finkbeiner},
  {Holmberg}, {Koppenhoefer}, {Price}, {Rest}, {Saglia}, {Schlafly}, {Smartt},
  {Sweeney}, {Wainscoat}, {Burgett}, {Chastel}, {Grav}, {Heasley}, {Hodapp},
  {Jedicke}, {Kaiser}, {Kudritzki}, {Luppino}, {Lupton}, {Monet}, {Morgan},
  {Onaka}, {Shiao}, {Stubbs}, {Tonry}, {White}, {Ba{\~n}ados}, {Bell},
  {Bender}, {Bernard}, {Boegner}, {Boffi}, {Botticella}, {Calamida},
  {Casertano}, {Chen}, {Chen}, {Cole}, {Deacon}, {Frenk}, {Fitzsimmons},
  {Gezari}, {Gibbs}, {Goessl}, {Goggia}, {Gourgue}, {Goldman}, {Grant},
  {Grebel}, {Hambly}, {Hasinger}, {Heavens}, {Heckman}, {Henderson}, {Henning},
  {Holman}, {Hopp}, {Ip}, {Isani}, {Jackson}, {Keyes}, {Koekemoer}, {Kotak},
  {Le}, {Liska}, {Long}, {Lucey}, {Liu}, {Martin}, {Masci}, {McLean}, {Mindel},
  {Misra}, {Morganson}, {Murphy}, {Obaika}, {Narayan}, {Nieto-Santisteban},
  {Norberg}, {Peacock}, {Pier}, {Postman}, {Primak}, {Rae}, {Rai}, {Riess},
  {Riffeser}, {Rix}, {R{\"o}ser}, {Russel}, {Rutz}, {Schilbach}, {Schultz},
  {Scolnic}, {Strolger}, {Szalay}, {Seitz}, {Small}, {Smith}, {Soderblom},
  {Taylor}, {Thomson}, {Taylor}, {Thakar}, {Thiel}, {Thilker}, {Unger},
  {Urata}, {Valenti}, {Wagner}, {Walder}, {Walter}, {Watters}, {Werner},
  {Wood-Vasey}, \& {Wyse}}]{Chambers2016}
{Chambers}, K.~C., {Magnier}, E.~A., {Metcalfe}, N., {et~al.} 2016, arXiv
  e-prints, arXiv:1612.05560.
\newblock \doarXiv{1612.05560}

\bibitem[{{Chandler} {et~al.}(2021){Chandler}, {Trujillo}, \&
  {Hsieh}}]{chandler2021_2005qn173}
{Chandler}, C.~O., {Trujillo}, C.~A., \& {Hsieh}, H.~H. 2021, \apjl, 922, L8,
  \dodoi{10.3847/2041-8213/ac365b}

\bibitem[{{Cochran} \& {Cochran}(2002)}]{Cochran2002}
{Cochran}, A.~L., \& {Cochran}, W.~D. 2002, \icarus, 157, 297,
  \dodoi{10.1006/icar.2002.6850}

\bibitem[{Craig {et~al.}(2021)Craig, Crawford, Seifert, Robitaille, Sipőcz,
  Walawender, Crawford, Vinícius, Ninan, Droettboom, Youn, Yash-10, Tollerud,
  Bowers, Bray, Bach, stottsco, Janga, walkerna22, Lim, Günther, Rol, A.,
  Bradley, Price-Whelan, Deil, Ryon, Lee, Barbary, \&
  Weiner}]{Craig2021-ccdproc2.2.0}
Craig, M., Crawford, S., Seifert, M., {et~al.} 2021, {astropy/ccdproc: 2.2.0 --
  Image combination performance}, 2.2.0,  Zenodo,
  \dodoi{10.5281/zenodo.4883632}

\bibitem[{{Di Sisto} \& {Rossignoli}(2020)}]{2020CeMDA.132...36D}
{Di Sisto}, R.~P., \& {Rossignoli}, N.~L. 2020, Celestial Mechanics and
  Dynamical Astronomy, 132, 36, \dodoi{10.1007/s10569-020-09971-7}

\bibitem[{{Dobson} {et~al.}(2021{\natexlab{a}}){Dobson}, {Fitzsimmons},
  {Schwamb}, {Kelley}, {Lister}, {Denneau}, {Heinze}, {Shingles}, {Smith},
  {Tonry}, {Weiland}, {Young}, {Benecchi}, \& {Verbiscer}}]{Dobson2021ATel}
{Dobson}, M., {Fitzsimmons}, A., {Schwamb}, M.~E., {et~al.} 2021{\natexlab{a}},
  The Astronomer's Telegram, 14903, 1

\bibitem[{{Dobson} {et~al.}(2021{\natexlab{b}}){Dobson}, {Schwamb},
  {Fitzsimmons}, {Kelley}, {Lister}, {Shingles}, {Denneau}, {Heinze}, {Smith},
  {Tonry}, {Weiland}, {Young}, {Benecchi}, \& {Verbiscer}}]{Dobson2021}
{Dobson}, M.~M., {Schwamb}, M.~E., {Fitzsimmons}, A., {et~al.}
  2021{\natexlab{b}}, Research Notes of the American Astronomical Society, 5,
  211, \dodoi{10.3847/2515-5172/ac26c9}

\bibitem[{{Duncan} \& {Levison}(1997)}]{1997Sci...276.1670D}
{Duncan}, M.~J., \& {Levison}, H.~F. 1997, Science, 276, 1670,
  \dodoi{10.1126/science.276.5319.1670}

\bibitem[{{Dybczy{\'n}ski} \& {Breiter}(2022)}]{Dybczynski2022}
{Dybczy{\'n}ski}, P.~A., \& {Breiter}, S. 2022, \aap, 657, A65,
  \dodoi{10.1051/0004-6361/202142227}

\bibitem[{{Dybczy{\'n}ski} \& {Kr{\'o}likowska}(2015)}]{Dybczynski2015}
{Dybczy{\'n}ski}, P.~A., \& {Kr{\'o}likowska}, M. 2015, \mnras, 448, 588,
  \dodoi{10.1093/mnras/stv013}

\bibitem[{{Eastman} {et~al.}(2014){Eastman}, {Brown}, {Hygelund}, {van Eyken},
  {Tufts}, \& {Barnes}}]{Eastman2014NRES}
{Eastman}, J.~D., {Brown}, T.~M., {Hygelund}, J., {et~al.} 2014, in \procspie,
  Vol. 9147, Ground-based and Airborne Instrumentation for Astronomy V (SPIE),
  914716, \dodoi{10.1117/12.2054699}

\bibitem[{{Farnham}(2021)}]{Farnham2021}
{Farnham}, T. 2021, The Astronomer's Telegram, 14759, 1

\bibitem[{{Farnham} {et~al.}(2000){Farnham}, {Schleicher}, \&
  {A'Hearn}}]{Farnham2000}
{Farnham}, T.~L., {Schleicher}, D.~G., \& {A'Hearn}, M.~F. 2000, \icarus, 147,
  180, \dodoi{10.1006/icar.2000.6420}

\bibitem[{{Farnham} {et~al.}(2007){Farnham}, {Wellnitz}, {Hampton}, {Li},
  {Sunshine}, {Groussin}, {McFadden}, {Crockett}, {A'Hearn}, {Belton},
  {Schultz}, \& {Lisse}}]{Farnham2007}
{Farnham}, T.~L., {Wellnitz}, D.~D., {Hampton}, D.~L., {et~al.} 2007, \icarus,
  187, 26, \dodoi{10.1016/j.icarus.2006.10.036}

\bibitem[{{Feaga} {et~al.}(2007){Feaga}, {A'Hearn}, {Sunshine}, {Groussin}, \&
  {Farnham}}]{Feaga2007}
{Feaga}, L.~M., {A'Hearn}, M.~F., {Sunshine}, J.~M., {Groussin}, O., \&
  {Farnham}, T.~L. 2007, \icarus, 190, 345,
  \dodoi{10.1016/j.icarus.2007.04.009}

\bibitem[{{Fitzsimmons} {et~al.}(2021){Fitzsimmons}, {Erasmus}, {Thirouin},
  {Hsieh}, \& {Green}}]{fitzsimmons2021_2005QN173}
{Fitzsimmons}, A., {Erasmus}, N., {Thirouin}, A., {Hsieh}, H.~H., \& {Green},
  D.~W.~E. 2021, Central Bureau Electronic Telegrams, 4995, 1

\bibitem[{{Fitzsimmons} {et~al.}(2020)}]{Fitzsimmons2020-P2020X1}
{Fitzsimmons}, A., {et~al.} 2020, CBET, 4895, 1

\bibitem[{{F{\"o}rster} {et~al.}(2021){F{\"o}rster}, {Cabrera-Vives},
  {Castillo-Navarrete}, {Est{\'e}vez}, {S{\'a}nchez-S{\'a}ez}, {Arredondo},
  {Bauer}, {Carrasco-Davis}, {Catelan}, {Elorrieta}, {Eyheramendy}, {Huijse},
  {Pignata}, {Reyes}, {Reyes}, {Rodr{\'\i}guez-Mancini}, {Ruz-Mieres},
  {Valenzuela}, {{\'A}lvarez-Maldonado}, {Astorga}, {Borissova}, {Clocchiatti},
  {De Cicco}, {Donoso-Oliva}, {Hern{\'a}ndez-Garc{\'\i}a}, {Graham},
  {Jord{\'a}n}, {Kurtev}, {Mahabal}, {Maureira}, {Mu{\~n}oz-Arancibia},
  {Molina-Ferreiro}, {Moya}, {Palma}, {P{\'e}rez-Carrasco}, {Protopapas},
  {Romero}, {Sabatini-Gacitua}, {S{\'a}nchez}, {San Mart{\'\i}n},
  {Sep{\'u}lveda-Cobo}, {Vera}, \& {Vergara}}]{Forster2021ALeRCE}
{F{\"o}rster}, F., {Cabrera-Vives}, G., {Castillo-Navarrete}, E., {et~al.}
  2021, \aj, 161, 242, \dodoi{10.3847/1538-3881/abe9bc}

\bibitem[{{Gaia Collaboration} {et~al.}(2021){Gaia Collaboration}, {Brown},
  {Vallenari}, {Prusti}, {de Bruijne}, {Babusiaux}, {Biermann}, {Creevey},
  {Evans}, {Eyer}, \& et~al.}]{GaiaEDR3}
{Gaia Collaboration}, {Brown}, A.~G.~A., {Vallenari}, A., {et~al.} 2021, \aap,
  649, A1, \dodoi{10.1051/0004-6361/202039657}

\bibitem[{Ginsburg {et~al.}(2019)Ginsburg, Sip{\H{o}}cz, Brasseur,
  Cowperthwaite, Craig, Deil, Guillochon, Guzman, Liedtke, Lim, Lockhart,
  Mommert, Morris, Norman, Parikh, Persson, Robitaille, Segovia, Singer,
  Tollerud, de~Val-Borro, Valtchanov, Woillez, \& the~astroquery
  collaboration}]{astroquery2019}
Ginsburg, A., Sip{\H{o}}cz, B.~M., Brasseur, C.~E., {et~al.} 2019, \aj, 157,
  98, \dodoi{10.3847/1538-3881/aafc33}

\bibitem[{{Giorgini} {et~al.}(1996){Giorgini}, {Yeomans}, {Chamberlin},
  {Chodas}, {Jacobson}, {Keesey}, {Lieske}, {Ostro}, {Standish}, \&
  {Wimberly}}]{Giorgini1996}
{Giorgini}, J.~D., {Yeomans}, D.~K., {Chamberlin}, A.~B., {et~al.} 1996, in
  AAS/Division for Planetary Sciences Meeting Abstracts \#28, 25.04

\bibitem[{{Guilbert-Lepoutre}(2012)}]{2012AJ....144...97G}
{Guilbert-Lepoutre}, A. 2012, \aj, 144, 97, \dodoi{10.1088/0004-6256/144/4/97}

\bibitem[{{Henden} {et~al.}(2016){Henden}, {Templeton}, {Terrell}, {Smith},
  {Levine}, \& {Welch}}]{Henden2016APASS}
{Henden}, A.~A., {Templeton}, M., {Terrell}, D., {et~al.} 2016, VizieR Online
  Data Catalog, II/336

\bibitem[{{Holt} {et~al.}(2021){Holt}, {Knight}, {Lister}, {Kelley}, {Ye},
  {Snodgrass}, {Opitom}, {Kokotanekova}, {Micheli}, \& {Schwamb}}]{Holt2021}
{Holt}, C.~E., {Knight}, M.~M., {Lister}, T., {et~al.} 2021, in AAS/Division
  for Planetary Sciences Meeting Abstracts, Vol.~53, AAS/Division for Planetary
  Sciences Meeting Abstracts, 210.14

\bibitem[{{Hsieh} {et~al.}(2004){Hsieh}, {Jewitt}, \&
  {Fern{\'a}ndez}}]{Hsieh2004}
{Hsieh}, H.~H., {Jewitt}, D.~C., \& {Fern{\'a}ndez}, Y.~R. 2004, \aj, 127,
  2997, \dodoi{10.1086/383208}

\bibitem[{{Hsieh} {et~al.}(2011){Hsieh}, {Meech}, \&
  {Pittichov{\'a}}}]{Hsieh2011}
{Hsieh}, H.~H., {Meech}, K.~J., \& {Pittichov{\'a}}, J. 2011, \apjl, 736, L18,
  \dodoi{10.1088/2041-8205/736/1/L18}

\bibitem[{{Hsieh} {et~al.}(2021){Hsieh}, {Chandler}, {Denneau}, {Fitzsimmons},
  {Erasmus}, {Kelley}, {Knight}, {Lister}, {Pittichov{\'a}}, {Sheppard},
  {Thirouin}, {Trujillo}, {Usher}, {Gomez}, {Chatelain}, {Greenstreet},
  {Angel}, {Miles}, {Roche}, \& {Wooding}}]{Hsieh2021_433P}
{Hsieh}, H.~H., {Chandler}, C.~O., {Denneau}, L., {et~al.} 2021, \apjl, 922,
  L9, \dodoi{10.3847/2041-8213/ac2c62}

\bibitem[{Huebner {et~al.}(2006)Huebner, Benkhoff, Capria, Coradini, Sanctis,
  Orosei, \& Prialnik}]{Huebner2006}
Huebner, W.~F., Benkhoff, J., Capria, M.~T., {et~al.} 2006, {Heat and Gas
  Diffusion in Comet Nuclei}, International Space Science Institute
  (International Space Science Institute)

\bibitem[{{Hughes}(1990)}]{Hughes1990}
{Hughes}, D.~W. 1990, \qjras, 31, 69

\bibitem[{{Ishiguro} {et~al.}(2016){Ishiguro}, {Kuroda}, {Hanayama}, {Kwon},
  {Kim}, {Lee}, {Watanabe}, {Akitaya}, {Kawabata}, {Itoh}, {Nakaoka},
  {Yoshida}, {Imai}, {Sarugaku}, {Yanagisawa}, {Ohta}, {Kawai}, {Miyaji},
  {Fukushima}, {Honda}, {Takahashi}, {Sato}, {Vaubaillon}, \&
  {Watanabe}}]{Ishiguro2016}
{Ishiguro}, M., {Kuroda}, D., {Hanayama}, H., {et~al.} 2016, \aj, 152, 169,
  \dodoi{10.3847/0004-6256/152/6/169}

\bibitem[{{Ivezi{\'c}} {et~al.}(2019){Ivezi{\'c}}, {Kahn}, {Tyson}, {Abel},
  {Acosta}, {Allsman}, {Alonso}, {AlSayyad}, {Anderson}, {Andrew}, \&
  et~al.}]{Ivezic2019}
{Ivezi{\'c}}, {\v Z}., {Kahn}, S.~M., {Tyson}, J.~A., {et~al.} 2019, \apj, 873,
  111, \dodoi{10.3847/1538-4357/ab042c}

\bibitem[{{Jacobson} {et~al.}(2014){Jacobson}, {Marzari}, {Rossi}, {Scheeres},
  \& {Davis}}]{Jacobson2014}
{Jacobson}, S.~A., {Marzari}, F., {Rossi}, A., {Scheeres}, D.~J., \& {Davis},
  D.~R. 2014, \mnras, 439, L95, \dodoi{10.1093/mnrasl/slu006}

\bibitem[{{Jehin} {et~al.}(2021){Jehin}, {Moulane}, \&
  {Manfroid}}]{Jehin2021ATel15128}
{Jehin}, E., {Moulane}, Y., \& {Manfroid}, J. 2021, The Astronomer's Telegram,
  15128, 1

\bibitem[{{Jehin} {et~al.}(2020){Jehin}, {Moulane}, {Manfroid}, {Pozuelos}, \&
  {Hutsemekers}}]{Jehin2020ATel14101}
{Jehin}, E., {Moulane}, Y., {Manfroid}, J., {Pozuelos}, F., \& {Hutsemekers},
  D. 2020, The Astronomer's Telegram, 14101, 1

\bibitem[{{Jehin} {et~al.}(2011){Jehin}, {Gillon}, {Queloz}, {Magain},
  {Manfroid}, {Chantry}, {Lendl}, {Hutsem{\'e}kers}, \& {Udry}}]{Jehin2011}
{Jehin}, E., {Gillon}, M., {Queloz}, D., {et~al.} 2011, The Messenger, 145, 2

\bibitem[{{Jewitt}(2009)}]{2009AJ....137.4296J}
{Jewitt}, D. 2009, \aj, 137, 4296, \dodoi{10.1088/0004-6256/137/5/4296}

\bibitem[{{Jewitt} {et~al.}(2014){Jewitt}, {Agarwal}, {Li}, {Weaver},
  {Mutchler}, \& {Larson}}]{Jewitt2014}
{Jewitt}, D., {Agarwal}, J., {Li}, J., {et~al.} 2014, \apjl, 784, L8,
  \dodoi{10.1088/2041-8205/784/1/L8}

\bibitem[{{Jewitt} {et~al.}(2015){Jewitt}, {Hsieh}, \& {Agarwal}}]{Jewitt2015}
{Jewitt}, D., {Hsieh}, H., \& {Agarwal}, J. 2015, in Asteroids IV (University
  of Arizona Press), 221--241,
  \dodoi{10.2458/azu\_uapress\_9780816532131-ch012}

\bibitem[{{Jewitt} {et~al.}(2011){Jewitt}, {Weaver}, {Mutchler}, {Larson}, \&
  {Agarwal}}]{Jewitt2011}
{Jewitt}, D., {Weaver}, H., {Mutchler}, M., {Larson}, S., \& {Agarwal}, J.
  2011, \apjl, 733, L4, \dodoi{10.1088/2041-8205/733/1/L4}

\bibitem[{{Jewitt}(2004)}]{Jewitt2004}
{Jewitt}, D.~C. 2004, in Comets II, ed. M.~C. {Festou}, H.~U. {Keller}, \&
  H.~A. {Weaver} (University of Arizona Press), 659

\bibitem[{{Joye} \& {Mandel}(2003)}]{Joye2003-DS9}
{Joye}, W.~A., \& {Mandel}, E. 2003, in Astronomical Society of the Pacific
  Conference Series, Vol. 295, Astronomical Data Analysis Software and Systems
  XII, ed. H.~E. {Payne}, R.~I. {Jedrzejewski}, \& R.~N. {Hook}, 489

\bibitem[{Kelley \& Lister(2021)}]{Kelley2021-calviacat}
Kelley, M., \& Lister, T. 2021, mkelley/calviacat: v1.2.0, v1.2.0,  Zenodo,
  \dodoi{10.5281/zenodo.5061298}

\bibitem[{{Kelley}(2021)}]{Kelley2021-120P}
{Kelley}, M. S.~P. 2021, The Astronomer's Telegram, 14876, 1

\bibitem[{{Kelley} {et~al.}(2021{\natexlab{a}}){Kelley}, Lin, Sharma, Kumar,
  Bhalerao, Anupama, \& Barway}]{Kelley2021-7P-b}
{Kelley}, M. S.~P., Lin, Z.-Y., Sharma, K., {et~al.} 2021{\natexlab{a}},
  Central Bureau Electronic Telegrams, 4997, 2

\bibitem[{{Kelley} \& {Lister}(2021)}]{Kelley2021-7P-a}
{Kelley}, M. S.~P., \& {Lister}, T. 2021, The Astronomer's Telegram, 14486, 1

\bibitem[{{Kelley} {et~al.}(2021{\natexlab{b}}){Kelley}, {Lister}, \&
  {Holt}}]{Kelley2021-BB}
{Kelley}, M. S.~P., {Lister}, T., \& {Holt}, C.~E. 2021{\natexlab{b}}, The
  Astronomer's Telegram, 14917, 1

\bibitem[{{Kelley} {et~al.}(2021{\natexlab{c}}){Kelley}, {Lister}, {Sharma},
  {Kumar}, {Bhalerao}, {Anupama}, \& {Barway}}]{Kelley2021-22P}
{Kelley}, M. S.~P., {Lister}, T., {Sharma}, K., {et~al.} 2021{\natexlab{c}},
  The Astronomer's Telegram, 14565, 1

\bibitem[{{Kelley} {et~al.}(2021{\natexlab{d}}){Kelley}, {Lister}, {Sharma},
  {Kumar}, {Bhalerao}, {Anupama}, \& {Barway}}]{Kelley2021-2020R4-a}
---. 2021{\natexlab{d}}, The Astronomer's Telegram, 14618, 1

\bibitem[{{Kelley} {et~al.}(2021{\natexlab{e}}){Kelley}, {Lister}, {Sharma},
  {Kumar}, {Bhalerao}, {Anupama}, \& {Barway}}]{Kelley2021-2020R4-b}
---. 2021{\natexlab{e}}, Central Bureau Electronic Telegrams, 4996, 1

\bibitem[{{Kelley} {et~al.}(2021{\natexlab{f}}){Kelley}, {Sharma}, {Bhalerao},
  {Anupama}, \& {Barway}}]{Kelley2021-99P}
{Kelley}, M. S.~P., {Sharma}, K., {Bhalerao}, V., {Anupama}, G.~C., \&
  {Barway}, S. 2021{\natexlab{f}}, The Astronomer's Telegram, 14628, 1

\bibitem[{{Kelley} {et~al.}(2021{\natexlab{g}}){Kelley}, {Sharma}, {Kumar},
  {Bhalerao}, {Anupama}, \& {Barway}}]{Kelley2021-44P}
{Kelley}, M. S.~P., {Sharma}, K., {Kumar}, H., {et~al.} 2021{\natexlab{g}}, The
  Astronomer's Telegram, 14787, 1

\bibitem[{{Kelley} {et~al.}(2019){Kelley}, {Bodewits}, {Ye}, {Laher}, {Masci},
  {Monkewitz}, {Riddle}, {Rusholme}, {Shupe}, \& {Soumagnac}}]{kelley2019}
{Kelley}, M. S.~P., {Bodewits}, D., {Ye}, Q., {et~al.} 2019, in Astronomical
  Society of the Pacific Conference Series, Vol. 523, Astronomical Data
  Analysis Software and Systems XXVII, ed. P.~J. {Teuben}, M.~W. {Pound}, B.~A.
  {Thomas}, \& E.~M. {Warner}, 471

\bibitem[{{Kelley} {et~al.}(2021{\natexlab{h}}){Kelley}, {Sharma}, {Swain},
  {Kumar}, {Bhalerao}, {Anupama}, {Barway}, Gardener, {Lister}, Usher, Angel,
  {Zwicky Transient Facility Collaboration}, {LCO Outbursting Objects Key
  Project}, \& {GROWTH India Collaboration}}]{Kelley2021-67P}
{Kelley}, M. S.~P., {Sharma}, K., {Swain}, V., {et~al.} 2021{\natexlab{h}}, The
  Astronomer's Telegram, 15053, 1.
\newblock \url{https://ui.adsabs.harvard.edu/abs/2021ATel15053....1K}

\bibitem[{{Kelley} {et~al.}(2021{\natexlab{i}}){Kelley}, {Lister}, {Sharma},
  {Swain}, {Kumar}, {Bhalerao}, {Anupama}, {Barway}, {Zwicky Transient Facility
  Collaboration}, {LCO Outbursting Objects Key Project}, \& {GROWTH India
  Collaboration}}]{Kelley2021-97P}
{Kelley}, M. S.~P., {Lister}, T., {Sharma}, K., {et~al.} 2021{\natexlab{i}},
  The Astronomer's Telegram, 15016, 1

\bibitem[{{Kelley} {et~al.}(2021{\natexlab{j}}){Kelley}, {Lister}, {Sharma},
  {Swain}, {Kumar}, {Bhalerao}, {Anupama}, {Barway}, {Zwicky Transient Facility
  Collaboration}, {LCO Outbursting Objects Key Project}, \& {GROWTH India
  Collaboration}}]{Kelley2021-191P}
---. 2021{\natexlab{j}}, The Astronomer's Telegram, 15052, 1.
\newblock \url{https://ui.adsabs.harvard.edu/abs/2021ATel15052....1K}

\bibitem[{{Kelley} {et~al.}(2021{\natexlab{k}}){Kelley}, {Lister}, {Sharma},
  {Swain}, {Kumar}, {Bhalerao}, {Anupama}, {Barway}, {Zwicky Transient Facility
  Collaboration}, {LCO Outbursting Objects Key Project}, \& {Growth India
  Collaboration}}]{Kelley2021-382P}
---. 2021{\natexlab{k}}, The Astronomer's Telegram, 14940, 1

\bibitem[{{Kelley} {et~al.}(2021{\natexlab{l}}){Kelley}, {Farnham}, {Li},
  {Bodewits}, {Snodgrass}, {Allen}, {Bellm}, {Coughlin}, {Drake}, {Duev},
  {Graham}, {Kupfer}, {Masci}, {Reiley}, {Walters}, {Dominik}, {J{\o}rgensen},
  {Andrews}, {Bach-M{\o}ller}, {Bozza}, {Burgdorf}, {Campbell-White}, {Dib},
  {Fujii}, {Hinse}, {Hundertmark}, {Khalouei}, {Longa-Pe{\~n}a}, {Rabus},
  {Rahvar}, {Sajadian}, {Skottfelt}, {Southworth}, {Tregloan-Reed},
  {Unda-Sanzana}, \& {Mindstep Collaboration}}]{Kelley2021-46P}
{Kelley}, M. S.~P., {Farnham}, T.~L., {Li}, J.-Y., {et~al.} 2021{\natexlab{l}},
  \psj, 2, 131, \dodoi{10.3847/PSJ/abfe11}

\bibitem[{{Knight} \& {Schleicher}(2011)}]{Knight2011}
{Knight}, M.~M., \& {Schleicher}, D.~G. 2011, \aj, 141, 183,
  \dodoi{10.1088/0004-6256/141/6/183}

\bibitem[{{Knight} \& {Schleicher}(2013)}]{Knight2013}
---. 2013, \icarus, 222, 691, \dodoi{10.1016/j.icarus.2012.06.004}

\bibitem[{{Knight} {et~al.}(2021){Knight}, {Schleicher}, \&
  {Farnham}}]{Knight2021}
{Knight}, M.~M., {Schleicher}, D.~G., \& {Farnham}, T.~L. 2021, \psj, 2, 104,
  \dodoi{10.3847/PSJ/abef6c}

\bibitem[{{Kokotanekova} {et~al.}(2021){Kokotanekova}, {Lister}, {Bannister},
  {Snodgrass}, {Opitom}, {Schwamb}, \& {Kelley}}]{Kokotanekova2021}
{Kokotanekova}, R., {Lister}, T., {Bannister}, M., {et~al.} 2021, The
  Astronomer's Telegram, 14733, 1

\bibitem[{{Kokotanekova} {et~al.}(2022){Kokotanekova}, Snodgrass, Knight, Holt,
  Bannister, Lister, Kelley, Opitom, Protopapa, Schwamb, Onken, \&
  Wolf}]{Kokotanekova2022}
{Kokotanekova}, R., Snodgrass, C., Knight, M.~M., {et~al.} 2022

\bibitem[{{Kronk}(2003)}]{Kronk2003-Cometography-Vol2}
{Kronk}, G.~W. 2003, {Cometography: A Catalog of Comets, Volume 2: 1800-1899}
  (Cambridge University Press)

\bibitem[{{Lauretta} {et~al.}(2019){Lauretta}, {Hergenrother}, {Chesley},
  {Leonard}, {Pelgrift}, {Adam}, {Al Asad}, {Antreasian}, {Ballouz}, {Becker},
  {Bennett}, {Bos}, {Bottke}, {Brozovi{\'c}}, {Campins}, {Connolly}, {Daly},
  {Davis}, {de Le{\'o}n}, {DellaGiustina}, {Drouet d'Aubigny}, {Dworkin},
  {Emery}, {Farnocchia}, {Glavin}, {Golish}, {Hartzell}, {Jacobson}, {Jawin},
  {Jenniskens}, {Kidd}, {Lessac-Chenen}, {Li}, {Libourel}, {Licandro},
  {Liounis}, {Maleszewski}, {Manzoni}, {May}, {McCarthy}, {McMahon}, {Michel},
  {Molaro}, {Moreau}, {Nelson}, {Owen}, {Rizk}, {Roper}, {Rozitis}, {Sahr},
  {Scheeres}, {Seabrook}, {Selznick}, {Takahashi}, {Thuillet}, {Tricarico},
  {Vokrouhlick{\'y}}, \& {Wolner}}]{Lauretta2019}
{Lauretta}, D.~S., {Hergenrother}, C.~W., {Chesley}, S.~R., {et~al.} 2019,
  Science, 366, 3544, \dodoi{10.1126/science.aay3544}

\bibitem[{{Le Roy} {et~al.}(2015){Le Roy}, {Altwegg}, {Balsiger}, {Berthelier},
  {Bieler}, {Briois}, {Calmonte}, {Combi}, {De Keyser}, {Dhooghe}, {Fiethe},
  {Fuselier}, {Gasc}, {Gombosi}, {H{\"a}ssig}, {J{\"a}ckel}, {Rubin}, \&
  {Tzou}}]{LeRoy2015}
{Le Roy}, L., {Altwegg}, K., {Balsiger}, H., {et~al.} 2015, \aap, 583, A1,
  \dodoi{10.1051/0004-6361/201526450}

\bibitem[{{Lellouch} {et~al.}(2022){Lellouch}, {Moreno}, {Bockel{\'e}e-Morvan},
  {Biver}, \& {Santos-Sanz}}]{Lellouch2022}
{Lellouch}, E., {Moreno}, R., {Bockel{\'e}e-Morvan}, D., {Biver}, N., \&
  {Santos-Sanz}, P. 2022, arXiv e-prints, arXiv:2201.13188.
\newblock \doarXiv{2201.13188}

\bibitem[{{Levison} {et~al.}(2009){Levison}, {Bottke}, {Gounelle},
  {Morbidelli}, {Nesvorn{\'y}}, \& {Tsiganis}}]{Levison2009}
{Levison}, H.~F., {Bottke}, W.~F., {Gounelle}, M., {et~al.} 2009, \nat, 460,
  364, \dodoi{10.1038/nature08094}

\bibitem[{{Li} {et~al.}(2011){Li}, {Jewitt}, {Clover}, \& {Jackson}}]{Li2011}
{Li}, J., {Jewitt}, D., {Clover}, J.~M., \& {Jackson}, B.~V. 2011, \apj, 728,
  31, \dodoi{10.1088/0004-637X/728/1/31}

\bibitem[{{Lister} {et~al.}(2021){Lister}, {Gomez}, {Chatelain}, {Greenstreet},
  {MacFarlane}, {Tedeschi}, \& {Kosic}}]{Lister2021NEOx}
{Lister}, T.~A., {Gomez}, E., {Chatelain}, J., {et~al.} 2021, \icarus, 364,
  114387, \dodoi{10.1016/j.icarus.2021.114387}

\bibitem[{{McCully} {et~al.}(2018){McCully}, {Volgenau}, {Harbeck}, {Lister},
  {Saunders}, {Turner}, {Siverd}, \& {Bowman}}]{McCully2018BANZAI}
{McCully}, C., {Volgenau}, N.~H., {Harbeck}, D.~R., {et~al.} 2018, in
  \procspie, Vol. 10707, Observatory Operations: Strategies, Processes, and
  Systems VII (SPIE), 107070K

\bibitem[{{McNaught}(2003)}]{McNaught2003-156P}
{McNaught}, R. 2003, IAUC, 8118, 1.
\newblock \url{http://www.cbat.eps.harvard.edu/iauc/08100/08118.html}

\bibitem[{{Meech} {et~al.}(2009){Meech}, {Pittichov{\'a}}, {Bar-Nun},
  {Notesco}, {Laufer}, {Hainaut}, {Lowry}, {Yeomans}, \& {Pitts}}]{Meech2009}
{Meech}, K.~J., {Pittichov{\'a}}, J., {Bar-Nun}, A., {et~al.} 2009, \icarus,
  201, 719, \dodoi{10.1016/j.icarus.2008.12.045}

\bibitem[{{Meech} {et~al.}(2017){Meech}, {Kleyna}, {Hainaut}, {Micheli},
  {Bauer}, {Denneau}, {Keane}, {Stephens}, {Jedicke}, {Wainscoat}, {Weryk},
  {Flewelling}, {Schunov{\'a}-Lilly}, {Magnier}, \& {Chambers}}]{Meech2017}
{Meech}, K.~J., {Kleyna}, J.~T., {Hainaut}, O., {et~al.} 2017, \apjl, 849, L8,
  \dodoi{10.3847/2041-8213/aa921f}

\bibitem[{Mommert {et~al.}(2019)Mommert, p.~Kelley, de~Val-Borro, Li, Guzman,
  Sipőcz, Ďurech, Granvik, Grundy, Moskovitz, Penttilä, \&
  Samarasinha}]{Mommert2019}
Mommert, M., p.~Kelley, M.~S., de~Val-Borro, M., {et~al.} 2019, Journal of Open
  Source Software, 4, 1426, \dodoi{10.21105/joss.01426}

\bibitem[{{Morbidelli} \& {Nesvorn{\'y}}(2020)}]{Morbidelli2020}
{Morbidelli}, A., \& {Nesvorn{\'y}}, D. 2020, in The Trans-Neptunian Solar
  System, ed. D.~{Prialnik}, M.~A. {Barucci}, \& L.~{Young} (Elsevier), 25--59,
  \dodoi{10.1016/B978-0-12-816490-7.00002-3}

\bibitem[{Moulane {et~al.}(2018)Moulane, Jehin, Opitom, Pozuelos, Manfroid,
  Benkhaldoun, Daassou, \& Gillon}]{Moulane2018}
Moulane, Y., Jehin, E., Opitom, C., {et~al.} 2018, \aap, 619, A156,
  \dodoi{10.1051/0004-6361/201833582}

\bibitem[{{Narayan} {et~al.}(2018){Narayan}, {Zaidi}, {Soraisam}, {Wang},
  {Lochner}, {Matheson}, {Saha}, {Yang}, {Zhao}, {Kececioglu}, {Scheidegger},
  {Snodgrass}, {Axelrod}, {Jenness}, {Maier}, {Ridgway}, {Seaman}, {Evans},
  {Singh}, {Taylor}, {Toeniskoetter}, {Welch}, {Zhu}, \& {ANTARES
  Collaboration}}]{Narayan2018}
{Narayan}, G., {Zaidi}, T., {Soraisam}, M.~D., {et~al.} 2018, \apjs, 236, 9,
  \dodoi{10.3847/1538-4365/aab781}

\bibitem[{{Narita} {et~al.}(2019){Narita}, {Fukui}, {Kusakabe}, {Watanabe},
  {Palle}, {Parviainen}, {Monta{\~n}{\'e}s-Rodr{\'\i}guez}, {Murgas},
  {Monelli}, {Aguiar}, {Perez Prieto}, {Oscoz}, {de Leon}, {Mori}, {Tamura},
  {Yamamuro}, {B{\'e}jar}, {Crouzet}, {Hidalgo}, {Klagyivik}, {Luque}, \&
  {Nishiumi}}]{Narita2019muscat2}
{Narita}, N., {Fukui}, A., {Kusakabe}, N., {et~al.} 2019, Journal of
  Astronomical Telescopes, Instruments, and Systems, 5, 015001,
  \dodoi{10.1117/1.JATIS.5.1.015001}

\bibitem[{{Narita} {et~al.}(2020){Narita}, {Fukui}, {Kusakabe}, {Watanabe},
  {Palle}, {Parviainen}, {Monta{\~n}{\'e}s-Rodr{\'\i}guez}, {Murgas},
  {Monelli}, {Aguiar}, {Perez Prieto}, {Oscoz}, {de Leon}, {Mori}, {Tamura},
  {Yamamuro}, {B{\'e}jar}, {Crouzet}, {Hidalgo}, {Klagyivik}, {Luque}, \&
  {Nishiumi}}]{Narita2020muscat3}
{Narita}, N., {Fukui}, A., {Kusakabe}, N., {et~al.} 2020, in Society of
  Photo-Optical Instrumentation Engineers (SPIE) Conference Series, Vol. 11447,
  Society of Photo-Optical Instrumentation Engineers (SPIE) Conference Series,
  114475K, \dodoi{10.1117/12.2559947}

\bibitem[{{Opitom} {et~al.}(2016){Opitom}, {Guilbert-Lepoutre}, {Jehin},
  {Manfroid}, {Hutsem{\'e}kers}, {Gillon}, {Magain}, {Roberts-Borsani}, \&
  {Witasse}}]{Opitom2016}
{Opitom}, C., {Guilbert-Lepoutre}, A., {Jehin}, E., {et~al.} 2016, \aap, 589,
  A8, \dodoi{10.1051/0004-6361/201527628}

\bibitem[{{Patterson} {et~al.}(2019){Patterson}, {Bellm}, {Rusholme}, {Masci},
  {Juric}, {Krughoff}, {Golkhou}, {Graham}, {Kulkarni}, {Helou}, \& {Zwicky
  Transient Facility Collaboration}}]{Patterson2019}
{Patterson}, M.~T., {Bellm}, E.~C., {Rusholme}, B., {et~al.} 2019, \pasp, 131,
  018001, \dodoi{10.1088/1538-3873/aae904}

\bibitem[{{Prialnik} \& {Rosenberg}(2009)}]{Prialnik2009}
{Prialnik}, D., \& {Rosenberg}, E.~D. 2009, \mnras, 399, L79,
  \dodoi{10.1111/j.1745-3933.2009.00727.x}

\bibitem[{{Ridden-Harper} {et~al.}(2021){Ridden-Harper}, {Bannister}, \&
  {Kokotanekova}}]{Ridden-Harper2021}
{Ridden-Harper}, R., {Bannister}, M.~T., \& {Kokotanekova}, R. 2021, Research
  Notes of the American Astronomical Society, 5, 161,
  \dodoi{10.3847/2515-5172/ac1512}

\bibitem[{Robitaille(2018)}]{Robitaille2018-reproject}
Robitaille, T. 2018, {reproject: astronomical image reprojection in Python},
  v0.4,  Zenodo, \dodoi{10.5281/zenodo.1162674}

\bibitem[{{Rossi} {et~al.}(2009){Rossi}, {Marzari}, \& {Scheeres}}]{Rossi2009}
{Rossi}, A., {Marzari}, F., \& {Scheeres}, D.~J. 2009, \icarus, 202, 95,
  \dodoi{10.1016/j.icarus.2009.02.030}

\bibitem[{Sako {et~al.}(2008)Sako, Sekiguchi, Sasaki, Okajima, Abe, Bond,
  Hearnshaw, Itow, Kamiya, Kilmartin, Masuda, Matsubara, Muraki, Rattenbury,
  Sullivan, Sumi, Tristram, Yanagisawa, \& Yock}]{Sako2008}
Sako, T., Sekiguchi, T., Sasaki, M., {et~al.} 2008, Experimental Astronomy, 22,
  51, \dodoi{10.1007/s10686-007-9082-5}

\bibitem[{{Schleicher} \& {Bair}(2011)}]{Schleicher2011}
{Schleicher}, D.~G., \& {Bair}, A.~N. 2011, \aj, 141, 177,
  \dodoi{10.1088/0004-6256/141/6/177}

\bibitem[{{Sharma} {et~al.}(2021{\natexlab{a}}){Sharma}, {Kelley}, {Stanzin},
  {Kumar}, {Bhalerao}, {Anupama}, \& {Barway}}]{Sharma2021-7p}
{Sharma}, K., {Kelley}, M. S.~P., {Stanzin}, J., {et~al.} 2021{\natexlab{a}},
  The Astronomer's Telegram, 14687, 1

\bibitem[{{Sharma} {et~al.}(2021{\natexlab{b}}){Sharma}, {Kelley}, {Lister},
  {Swain}, {Kumar}, {Bhalerao}, {Anupama}, {Growth-India}, {Zwicky Transient
  Collaboration}, \& {Lco Outbursting Objects Key Project
  Collaboration}}]{Sharma2021-29P}
{Sharma}, K., {Kelley}, M. S.~P., {Lister}, T., {et~al.} 2021{\natexlab{b}},
  The Astronomer's Telegram, 14984, 1

\bibitem[{{Siverd} {et~al.}(2018){Siverd}, {Brown}, {Barnes}, {Bowman}, {De
  Vera}, {Foale}, {Harbeck}, {Henderson}, {Hygelund}, {Kirby}, {McCully},
  {Nation}, {Smith}, {Taylor}, \& {Tufts}}]{Siverd2018NRES}
{Siverd}, R.~J., {Brown}, T.~M., {Barnes}, S., {et~al.} 2018, in Society of
  Photo-Optical Instrumentation Engineers (SPIE) Conference Series, Vol. 10702,
  \procspie, 107026C, \dodoi{10.1117/12.2312800}

\bibitem[{{Smith} {et~al.}(2019){Smith}, {Williams}, {Young}, {Ibsen},
  {Smartt}, {Lawrence}, {Morris}, {Voutsinas}, \& {Nicholl}}]{Smith2019Lasair}
{Smith}, K.~W., {Williams}, R.~D., {Young}, D.~R., {et~al.} 2019, Research
  Notes of the American Astronomical Society, 3, 26,
  \dodoi{10.3847/2515-5172/ab020f}

\bibitem[{{Snodgrass} \& {Jones}(2019)}]{Snodgrass2019}
{Snodgrass}, C., \& {Jones}, G.~H. 2019, Nature Communications, 10, 5418,
  \dodoi{10.1038/s41467-019-13470-1}

\bibitem[{{Snodgrass} {et~al.}(2017){Snodgrass}, {Agarwal}, {Combi},
  {Fitzsimmons}, {Guilbert-Lepoutre}, {Hsieh}, {Hui}, {Jehin}, {Kelley},
  {Knight}, {Opitom}, {Orosei}, {de Val-Borro}, \& {Yang}}]{snodgrass2017}
{Snodgrass}, C., {Agarwal}, J., {Combi}, M., {et~al.} 2017, \aapr, 25, 5,
  \dodoi{10.1007/s00159-017-0104-7}

\bibitem[{Stern(2003)}]{Stern2003}
Stern, S.~A. 2003, Nature, 424, 639, \dodoi{10.1038/nature01725}

\bibitem[{{Street} {et~al.}(2020){Street}, {Adamson}, {Blakeslee}, {Blum},
  {Bolton}, {Boroson}, {Bowman}, {Brice{\~n}o}, {Elias}, {Gomez}, {Heathcote},
  {Heinrich-Josties}, {Hopkinson}, {Lee}, {Miller}, {Nation}, {Ridgway},
  {Silva}, \& {Storrie-Lombardi}}]{Street2020AEON}
{Street}, R.~A., {Adamson}, A., {Blakeslee}, J.~P., {et~al.} 2020, in Society
  of Photo-Optical Instrumentation Engineers (SPIE) Conference Series, Vol.
  11449, Observatory Operations: Strategies, Processes, and Systems VIII,
  1144925, \dodoi{10.1117/12.2559986}

\bibitem[{{Tonry} {et~al.}(2012){Tonry}, {Stubbs}, {Lykke}, {Doherty},
  {Shivvers}, {Burgett}, {Chambers}, {Hodapp}, {Kaiser}, {Kudritzki},
  {Magnier}, {Morgan}, {Price}, \& {Wainscoat}}]{Tonry2012PS1}
{Tonry}, J.~L., {Stubbs}, C.~W., {Lykke}, K.~R., {et~al.} 2012, \apj, 750, 99,
  \dodoi{10.1088/0004-637X/750/2/99}

\bibitem[{{Tonry} {et~al.}(2018{\natexlab{a}}){Tonry}, {Denneau}, {Heinze},
  {Stalder}, {Smith}, {Smartt}, {Stubbs}, {Weiland}, \&
  {Rest}}]{Tonry2018ATLAS}
{Tonry}, J.~L., {Denneau}, L., {Heinze}, A.~N., {et~al.} 2018{\natexlab{a}},
  \pasp, 130, 064505

\bibitem[{{Tonry} {et~al.}(2018{\natexlab{b}}){Tonry}, {Denneau}, {Flewelling},
  {Heinze}, {Onken}, {Smartt}, {Stalder}, {Weiland}, \&
  {Wolf}}]{Tonry2018refcat2}
{Tonry}, J.~L., {Denneau}, L., {Flewelling}, H., {et~al.} 2018{\natexlab{b}},
  \apj, 867, 105, \dodoi{10.3847/1538-4357/aae386}

\bibitem[{{Trigo-Rodr{\'\i}guez} {et~al.}(2008){Trigo-Rodr{\'\i}guez},
  {Garc{\'\i}a-Melendo}, {Davidsson}, {S{\'a}nchez}, {Rodr{\'\i}guez},
  {Lacruz}, {de Los Reyes}, \& {Pastor}}]{Trigo-Rodriguez2008}
{Trigo-Rodr{\'\i}guez}, J.~M., {Garc{\'\i}a-Melendo}, E., {Davidsson},
  B.~J.~R., {et~al.} 2008, \aap, 485, 599, \dodoi{10.1051/0004-6361:20078666}

\bibitem[{{Tseng} {et~al.}(2007){Tseng}, {Bockel{\'e}e-Morvan}, {Crovisier},
  {Colom}, \& {Ip}}]{Tseng2007}
{Tseng}, W.~L., {Bockel{\'e}e-Morvan}, D., {Crovisier}, J., {Colom}, P., \&
  {Ip}, W.~H. 2007, \aap, 467, 729, \dodoi{10.1051/0004-6361:20066666}

\bibitem[{{van Buitenen}(2021)}]{vanBuitenen2021-7P}
{van Buitenen}, G. 2021, Central Bureau Electronic Telegrams, 4997, 1

\bibitem[{{Veverka} {et~al.}(2013){Veverka}, {Klaasen}, {A'Hearn}, {Belton},
  {Brownlee}, {Chesley}, {Clark}, {Economou}, {Farquhar}, {Green}, {Groussin},
  {Harris}, {Kissel}, {Li}, {Meech}, {Melosh}, {Richardson}, {Schultz},
  {Silen}, {Sunshine}, {Thomas}, {Bhaskaran}, {Bodewits}, {Carcich},
  {Cheuvront}, {Farnham}, {Sackett}, {Wellnitz}, \& {Wolf}}]{Veverka2013}
{Veverka}, J., {Klaasen}, K., {A'Hearn}, M., {et~al.} 2013, \icarus, 222, 424,
  \dodoi{10.1016/j.icarus.2012.03.034}

\bibitem[{{Vincent} {et~al.}(2016){Vincent}, {A'Hearn}, {Lin}, {El-Maarry},
  {Pajola}, {Sierks}, {Barbieri}, {Lamy}, {Rodrigo}, {Koschny}, {Rickman},
  {Keller}, {Agarwal}, {Barucci}, {Bertaux}, {Bertini}, {Besse}, {Bodewits},
  {Cremonese}, {Da Deppo}, {Davidsson}, {Debei}, {De Cecco}, {Deller},
  {Fornasier}, {Fulle}, {Gicquel}, {Groussin}, {Guti{\'e}rrez},
  {Guti{\'e}rrez-Marquez}, {G{\"u}ttler}, {H{\"o}fner}, {Hofmann}, {Hviid},
  {Ip}, {Jorda}, {Knollenberg}, {Kovacs}, {Kramm}, {K{\"u}hrt}, {K{\"u}ppers},
  {Lara}, {Lazzarin}, {Lopez Moreno}, {Marzari}, {Massironi}, {Mottola},
  {Naletto}, {Oklay}, {Preusker}, {Scholten}, {Shi}, {Thomas}, {Toth}, \&
  {Tubiana}}]{Vincent2016}
{Vincent}, J.~B., {A'Hearn}, M.~F., {Lin}, Z.~Y., {et~al.} 2016, \mnras, 462,
  S184, \dodoi{10.1093/mnras/stw2409}

\bibitem[{{Vincent} {et~al.}(2017){Vincent}, {Hviid}, {Mottola}, {Kuehrt},
  {Preusker}, {Scholten}, {Keller}, {Oklay}, {de Niem}, {Davidsson}, {Fulle},
  {Pajola}, {Hofmann}, {Hu}, {Rickman}, {Lin}, {Feller}, {Gicquel},
  {Boudreault}, {Sierks}, {Barbieri}, {Lamy}, {Rodrigo}, {Koschny}, {A'Hearn},
  {Barucci}, {Bertaux}, {Bertini}, {Cremonese}, {Da Deppo}, {Debei}, {De
  Cecco}, {Deller}, {Fornasier}, {Groussin}, {Guti{\'e}rrez},
  {Guti{\'e}rrez-Marquez}, {G{\"u}ttler}, {Ip}, {Jorda}, {Knollenberg},
  {Kovacs}, {Kramm}, {K{\"u}ppers}, {Lara}, {Lazzarin}, {Lopez Moreno},
  {Marzari}, {Naletto}, {Penasa}, {Shi}, {Thomas}, {Toth}, \&
  {Tubiana}}]{Vincent2017}
{Vincent}, J.~B., {Hviid}, S.~F., {Mottola}, S., {et~al.} 2017, \mnras, 469,
  S329, \dodoi{10.1093/mnras/stx1691}

\bibitem[{{Walsh} {et~al.}(2011){Walsh}, {Morbidelli}, {Raymond}, {O'Brien}, \&
  {Mandell}}]{Walsh2011}
{Walsh}, K.~J., {Morbidelli}, A., {Raymond}, S.~N., {O'Brien}, D.~P., \&
  {Mandell}, A.~M. 2011, \nat, 475, 206, \dodoi{10.1038/nature10201}

\bibitem[{{Whipple}(1978)}]{Whipple1978}
{Whipple}, F.~L. 1978, Moon and Planets, 18, 343, \dodoi{10.1007/BF00896489}

\bibitem[{{Willmer}(2018)}]{Willmer2018}
{Willmer}, C. N.~A. 2018, \apjs, 236, 47, \dodoi{10.3847/1538-4365/aabfdf}

\bibitem[{{Yang} {et~al.}(2021){Yang}, {Jewitt}, {Zhao}, {Jiang}, {Ye}, \&
  {Chen}}]{Yang2021}
{Yang}, B., {Jewitt}, D., {Zhao}, Y., {et~al.} 2021, \apjl, 914, L17,
  \dodoi{10.3847/2041-8213/ac03b7}

\bibitem[{{Ye} {et~al.}(2019){Ye}, {Kelley}, {Bodewits}, {Bolin}, {Jones},
  {Lin}, {Bellm}, {Dekany}, {Duev}, {Groom}, {Helou}, {Kulkarni}, {Kupfer},
  {Masci}, {Prince}, \& {Soumagnac}}]{Ye2019}
{Ye}, Q., {Kelley}, M. S.~P., {Bodewits}, D., {et~al.} 2019, \apjl, 874, L16,
  \dodoi{10.3847/2041-8213/ab0f3c}

\end{thebibliography}
\bibliographystyle{aasjournal}

\allauthors

\end{CJK*}

\end{document}